\documentclass[12pt,english,floatfix,nofootinbib,superscriptaddress,aps,prd,preprint]{revtex4}
\usepackage[utf8]{inputenc}
\usepackage{float}
\usepackage{array}
\usepackage{bbold}
\usepackage{lipsum}
\usepackage{dsfont}
\usepackage{graphicx}
\usepackage{amsmath,amsthm,amsfonts,amssymb}
\usepackage{graphicx}
\usepackage[english]{babel} 
\usepackage{color}
\usepackage{tensor}
\usepackage{esint}
\usepackage[dvips]{epsfig}
\usepackage[dvips]{graphicx}
\usepackage{float}
\usepackage{units}
\usepackage{textcomp}
\usepackage{mathrsfs}
\usepackage{amsmath}
\usepackage[makeroom]{cancel}
\usepackage{amssymb}
\usepackage{amsbsy}
\usepackage{amsfonts}
\usepackage{amssymb,mathrsfs,xcolor}
\usepackage{esint}
\usepackage{braket}
\usepackage{array}
\usepackage{graphicx}

\usepackage{wasysym}
\usepackage{multirow}
\usepackage{wrapfig}
\usepackage{subfig}

\usepackage{stmaryrd}
\usepackage{upgreek}

\makeatletter

\makeatletter\usepackage{babel}

\usepackage{hyperref}
\hypersetup{
    colorlinks,
    citecolor=blue,
    filecolor=green,
    linkcolor=purple,
    urlcolor=red,
}

\usepackage{slashed}

\newcommand{\ie}{\begin{equation}}
\newcommand{\fe}{\end{equation}}
\newcommand{\se}{\begin{eqnarray}}
\newcommand{\ff}{\end{eqnarray}}

\begin{document}

\title{Thermodynamics of a quantum ring modified by Lorentz violation}

\author{A. A. Ara\'{u}jo Filho}
\email{dilto@fisica.ufc.br}

\affiliation{Departamento de Física Teórica and IFIC, Centro Mixto Universidad de Valencia--CSIC. Universidad
de Valencia, Burjassot-46100, Valencia, Spain}

\author{H. Hassanabadi}
\email{hha1349@gmail.com}

\affiliation{Physics Department, Shahrood University of Technology, Shahrood, Iran}

\author{J. A. A. S. Reis}
\email{jalfieres@gmail.com}

\affiliation{Programa de Pós-graduação em Física, Universidade Federal do Maranh\~{a}o (UFMA),\\ Campus Universit\'{a}rio do Bacanga, S\~{a}o Lu\'{\i}s -- MA, 65080-805, -- Brazil}

\affiliation{Universidade Estadual do Sudoeste da Bahia (UESB), Departamento de Ciências Exatas e Naturais, Campus Juvino Oliveira, Itapetinga -- BA, 45700-00,--Brazil}
 
\author{L. Lisboa-Santos}
\email{let\_lisboa@hotmail.com} 

\affiliation{Programa de Pós-graduação em Física, Universidade Federal do Maranh\~{a}o (UFMA),\\ Campus Universit\'{a}rio do Bacanga, S\~{a}o Lu\'{\i}s -- MA, 65080-805, -- Brazil}



\date{\today}

\begin{abstract}

In this work, we investigate the consequences of Lorentz-violating terms in the thermodynamic properties of a 1-dimensional quantum ring. In particular, we use the ensemble theory to obtain our results of interest. The thermodynamic functions as well as the spin currents are calculated as a function of the temperature. We observe that parameter $\xi$, which triggers the Lorentz symmetry breaking, plays a major role in low temperature regime. Finally, depending on the configuration of the system, electrons can rotate in two different directions: clockwise and counterclockwise.

\end{abstract}

\maketitle


\section{Introduction}

The standard model extension (SME) \cite{1,1.2,colladay1998lorentz,kostelecky2002signals,kostelecky2004gravity,kostelecky1989spontaneous} was created as an enlargement of the minimal standard model, possessing tensors which trigger the Lorentz symmetry breaking. Particularly, in its gauge sector, there exist much investigation in a variety of contexts \cite{2,3,4,5,6,11,21,31,41,51,61}, involving interactions of photon-fermion \cite{7}, nonminimal couplings \cite{8}, and operators of higher mass dimension \cite{9}. Investigations focusing on the fermionic sector of the SME were correlated to the violation of CPT symmetry \cite{10}, which was a productive environment for the subsequent applications \cite{11,12,13,13.2,12.2}.

Condensed matter systems represent a fruitful field for studying Lorentz symmetry breaking \cite{kostelecky2013fermions,reis2021thermal}, because there can exist privileged spacetime directions due to the tensor fields, leading a breakdown of rotation invariance. Two dimensional electronic systems form a remarkable environment worthy to be explored \cite{winkler2003spin}, where spin-orbit interaction has a relevant role, since there exists a correlation with the spintronics \cite{bliokh2015spin}. In this case, persistent currents and geometrical phases turn out to be observables.

A relevant feature when we deal with Lorentz violation, encoded by modified dispersion relations, is definitely its thermodynamic aspects. Studies concerning thermodynamic properties in Lorentz--violating scenarios could supply additional information about the primordial stages of the Universe \cite{kostelecky2011data}.
The correlation between them was initially addressed in Ref. \cite{colladay2004statistical}. After that, many works have been proposed in the literature within different contexts, namely, Podolsky electrodynamics \cite{araujo2021thermodynamic} and electrodynamics with an additive Myers--Pospelov term \cite{anacleto2018lorentz,myers2003ultraviolet,mariz2012perturbative}, graviton \cite{aa2021lorentz}, CPT--even and CPT--odd violations \cite{casana2008lorentz,casana2009finite,araujo2021higher}, higher-dimensional operators \cite{reis2021thermal}, Einstein--eather theory \cite{aaa2021thermodynamics}, bouncing universe \cite{petrov2021bouncing}, and loop quantum gravity \cite{aa2022particles}.

There are several works involving the study of quantum rings within different scenarios \cite{khordad2018thermodynamic,oliveira2019thermodynamic,chakraborty1994electron,warburton2000optical,fuhrer2001energy,bakke2014lorentz,bakke2015influence}. Additionally, a recent study appeared in the literature, focusing on the main classical aspects of 1-D quantum rings with Lorentz violation \cite{Manoel2015GeometricPhase}. Nevertheless, up to now, there is a lack in the literature concerning an investigation of the thermodynamic properties of quantum rings in this latter context. Such examination is notable since it might possibly reveal different phenomena ascribed to a new physics, which might be detected in a future laboratory experiment. It is important to mention that, throughout of this manuscript, natural units have been used. This means, for instance, that the Boltzmann constant is set to $\mathrm{k}_{B}=1$.


\section{Thermodynamic description}

Initially, we regard the general quadratic sector of a renormalizable CPT--violating Lagrangian, which describes a massive Dirac fermion
\ie
\mathcal{L} = \frac{i}{2} \Bar{\psi} \Gamma^{\nu}\overset{\leftrightarrow}{\partial}_{\nu}\psi - \Bar{\psi}M\psi \label{Lagrangian},
\fe
with $\Gamma^{\nu}:= \gamma^{\nu} + c^{\mu\nu}\gamma_{\mu} + d^{\mu\nu}\gamma_{5}\gamma_{\mu} + e^{\nu} + i f^{\nu}\gamma_{5} + g^{\gamma\mu\nu}\sigma_{\lambda\mu}/2$ and $M := m + a_{\mu}\gamma^{\mu} + b_{\mu}\gamma_{5}\gamma^{\mu} + H^{\mu\nu}\sigma_{\mu\nu}/2$, where the gamma matrices have the conventional properties, i.e., $\mathbb{1}, \gamma_{5}, \gamma^{\mu}, \gamma_{5}\gamma^{\mu}, \sigma^{\mu\nu}$. In addition, parameters $a_{\mu}, b_{\mu}, c_{\mu\nu}, H_{\mu\nu}$ are given by the vacuum expectation values of Lorentz tensors, which come from the spontaneous Lorentz--breaking of a more fundamental theory \cite{kostelecky2001stability}. More so, the Hamiltonian from Eq. (\ref{Lagrangian}) can also be derived as
\ie
H = - A^{\dagger}\gamma^{0}(i \Gamma^{j}\partial_{j}- M)A
\fe
where $A$ is a non--singular matrix(not depending on the spacetime), which satisfy the condition below
\ie
A^{\dagger}\gamma^{0}\Gamma^{0}A = \mathbb{1}.
\fe
Now, let us consider the recent results in literature proposed by Ref. \cite{Manoel2015GeometricPhase}. Mainly, there exists an effective nonrelativistic Hamiltonian
\begin{equation}
H=\frac{p^{2}}{2m}+d_{jk}p^{j}\sigma ^{k}+d_{00}p^{i}\sigma ^{i},
\end{equation}%
which provides structures analogous to Rashba and Dresselhauss interactions. When the system is restricted to a 1--dimensional quantum ring, two main features give rise to: the appearance of geometrical phases and persistent spin currents effects. Fundamentally, they occur due to the operators $d_{jk}$ and $d_{00}$. The respective energy eigenvalues are obtained as follows:%
\begin{equation}
E_{n,s}^{d_{jk}}=\Omega \left\{ n-\frac{1}{2}\left[ 1+\left( -1\right) ^{s}%
\sqrt{1+4\xi ^{2}}\right] \right\} ^{2},  \label{eq:Dij-case-energy}
\end{equation}%
\begin{equation}
E_{n,s}^{d_{00}}=\Omega \left\{ n+\frac{1}{2}\left[ 1-\left( -1\right) ^{s}%
\sqrt{1+4\xi _{00}^{2}}\right] \right\} ^{2},  \label{eq:D00-case-energy}
\end{equation}%
where $n\in \mathbb{Z}$, $\Omega=1/2\mathrm{m}r_{0}^{2}$, $s=1$ represents the energy for spin-down particles while
$s=2$ represents the energy for spin-up ones. The information about Lorentz violation relies on
the parameters $\xi=\mathrm{m}r_{0}\sqrt{d_{12}^{2}+d_{11}^{2}} $ and $\xi _{00}=\mathrm{m}r_{0}d_{00}$, where $\mathrm{m}$ is the electron mass and $r_{0}$ is the radius of the quantum ring. It is important to mention that we have used $p_{z}=0$, since electrons move on the plane. We have also set $d^{32}=d^{31}=d^{33}=0$, so that $d^{ij}$ becomes a $2\times2$ matrix. Choosing $d^{22}=-d^{11}$, tensor $d^{\mu\nu}$ becomes traceless, as required. Also, the components $d^{0i}$ and $d^{0i}$ are set to be zero because they do not fit with our purpose, the Rashba- and Dresselhauss-like interactions. From now on, we shall describe the properties of interest based on parameters $\xi$ and $\xi_{00}$, instead of purely Lorentz--violating parameters themselves. We choose such a configuration due to the necessity of keeping both mass and radius fixed in order to vary the controlling coefficients only. It is worthy to be mentioned that the eigenstates were also
calculated in Ref. \cite{Manoel2015GeometricPhase}. Nevertheless, only with the spectral energy our main results of interest can be obtained. The spin current may be calculated for an arbitrary state as \cite{Manoel2015GeometricPhase}
\begin{equation}
\mathcal{J}_{\varphi }^{z}=\frac{1}{4\mathrm{m}r_{0}}\left( 2n\cos \theta -1\right) ,
\label{eq:spinCurrent}
\end{equation}%
where $\theta$ can assume the form of either $ \theta = \arctan (2\xi) $ or $ \theta = \arctan(2\xi_{00}$), i.e., if we consider $d_{jk}$ or $d_{00}$ respectively. In our work, we investigate the consequences of Lorentz-violating dispersion relations present in Eqs.
$\left( \ref{eq:Dij-case-energy}\right) $ and $\left( \ref%
{eq:D00-case-energy}\right)$ within the thermodynamic properties of a 1-dimensional quantum ring. To do so,
we apply two different methods to study the thermal aspects of non-interacting electrons of our system.

The first method is based on the canonical
ensemble formalism. On the other hand, the second one is founded on the grand canonical ensemble theory. These approaches allow us to study $N$ electrons confined to a 1-dimensional quantum ring in contact with a thermal reservoir. 
Additionally, it is worth pointing out that using the first method, we are not able to take into account the Fermi-Dirac statistics since there does not exist the differentiation ascribed to the quantum number. Then, electrons are free to choose its accessible quantum states. However, for the second situation, we are able to take into account the Fermi-Dirac statistics.

In the canonical ensemble, it is well-known that the partition function is given by%
\begin{equation}
\mathcal{Z}=\sum_{\left\{ n_{j}\right\} }\exp \left( -\beta E_{\left\{
n_{j}\right\} }\right) ,  \label{eq:Partition-function}
\end{equation}%
where $\left\{ n_{j}\right\} $ is related to all accessible quantum states. Since we are dealing with $N$ non-interacting particles, the partition function in Eq. $%
\left( \ref{eq:Partition-function}\right) $ can be factorized \cite{araujo2022fermions,oliveira2020relativistic}%
\begin{equation}
\mathcal{Z}=\mathcal{Z}_{1}^{N}=\left\{ \sum_{n=0}^{\infty
}\sum_{s=1}^{2}\exp \left( -\beta E_{n,s}\right) \right\} ^{N},
\label{eq:Partition-function-1}
\end{equation}%
where we have defined the single partition function as
\begin{equation}
\mathcal{Z}_{1}=\sum_{n=0}^{\infty }\sum_{s=1}^{2}\exp \left( -\beta
E_{n,s}\right) .  \label{eq:Single-Partition-function}
\end{equation}%
In order to obtain an analytical expression for it, we use the \textit{Euler--Maclaurin} formula \cite{oliveira2020relativistic,oliveira2020reply,oliveira2020thermodynamic,kac2002euler}
\begin{align}
\sum_{n=0}^{\infty }F\left( n\right) & =\int_{0}^{\infty }F\left( n\right) d%
\mathrm{n}+\frac{1}{2}F\left( 0\right) \displaybreak[0]  \notag \\
& -\frac{1}{2!}B_{2}F^{\prime }\left( 0\right) -\frac{1}{4!}B_{4}F^{\prime
\prime \prime }\left( 0\right) +\ldots ,
\end{align}%
where $B_{m}$ are the Bernoulli numbers: $B_{2}=1/6$, $B_{4}=-1/30$, $\ldots
$. Performing the calculations, we get%
\begin{eqnarray}
\mathcal{Z}_{1}^{d_{jk}} &=&\sqrt{\frac{\pi }{\beta \Omega }}\left\{ 1+\frac{%
1}{2}\sum_{s=1}^{2}\mathrm{Erf}\left[ \sqrt{\frac{\beta \Omega }{4}}\Delta
_{s}\right] \right\} +  \notag \\
&&\sum_{s=1}^{2}\exp \left( -\frac{\beta \Omega }{4}\Delta _{s}^{2}\right)
\times   \notag \\
&&\times \left\{ \frac{1}{2}-\frac{\beta \Omega }{12}\left( 1+\frac{\beta
\Omega }{10}\right) \Delta _{s}+\frac{\beta ^{3}\Omega ^{3}}{720}\Delta
_{s}^{3}+\ldots \right\} \label{eq:ZfunctionDij},
\end{eqnarray}
where we have defined%
\begin{eqnarray}
\Delta _{s} &\equiv &1+\left( -1\right) ^{s}\sqrt{1+4\xi ^{2}}, \\
\mathrm{Erf}\left[ z\right]  &=&\frac{2}{\sqrt{\pi }}%
\int_{0}^{z}e^{-t^{2}}dt.
\end{eqnarray}%
In particular, for the coefficient $d_{00}$, we find%
\begin{eqnarray}
\mathcal{Z}_{1}^{d_{00}} &=&\frac{1}{2}\sqrt{\frac{\pi }{\beta \Omega }}%
\sum_{s=1}^{2}\mathrm{Erf}\left[ \sqrt{\frac{\beta \Omega }{4}}\Psi _{s}%
\right] +  \notag \\
&&\sum_{s=1}^{2}\exp \left( -\frac{\beta \Omega }{4}\Psi _{s}^{2}\right)
\times   \notag \\
&&\times \left\{ \frac{1}{2}+\frac{\beta \Omega }{12}\left( 1+\frac{\beta
\Omega }{10}\right) \Psi _{s}-\frac{\beta ^{3}\Omega ^{3}}{720}\Psi
_{s}^{3}+\ldots \right\} \label{eq:ZfunctionD00} ,
\end{eqnarray}%
where%
\begin{equation}
\Psi _{s}\equiv 1-\left( -1\right) ^{s}\sqrt{1+4\xi _{00}^{2}}.
\end{equation}
Here, we must take care about the convergence of the power series. In both Eq. (\ref{eq:ZfunctionDij}) and (\ref{eq:ZfunctionD00}), we can see the presence of an exponential in front of the power series. Such term is a function of temperature and, for the range that we choose to evaluate the thermodynamic functions, $0<T<1~\mathrm{eV}$, it is possible to ensure the convergence of the partition functions and their respective thermodynamic quantities. Nevertheless, for a different range of temperatures, an additional study about the convergence must be carried out. It is important to mention that this examination lies beyond of the scope of this current work.

Moreover, the connection with the thermodynamics is given by the
Helmholtz free energy%
\begin{equation}
f=-\frac{1}{\beta }\lim_{N\rightarrow \infty }\frac{1}{N}\ln \mathcal{Z},
\end{equation}%
where we write rather the Helmholtz free energy per particle. With this, we can derive the following thermodynamic state functions, i.e., entropy, heat capacity and mean energy respectively,%
\begin{equation}
s=-\frac{\partial f}{\partial T},\phantom{a}c=T\frac{\partial s}{\partial T},%
\phantom{a}u=T^{2}\frac{\partial }{\partial T}\ln \mathcal{Z}.
\end{equation}%
For the grand canonical ensemble, the grand partition function reads%
\begin{equation}
\Xi =\sum_{N=0}^{\infty }\exp \left( \beta \mu N\right) \mathcal{Z}\left[
N_{n,s}\right] ,  \label{eq:GarndPartition-function}
\end{equation}%
where $\mathcal{Z}\left[ N_{n,s}\right] $ is the usual canonical partition
function. Since we are dealing with fermions, the
occupation number allowed for each quantum state is restricted to $%
N_{n,s}=\left\{ 0,1\right\} $. So, for an arbitrary quantum state, the
energy also depends of the occupation number as
\begin{equation}
E\left\{ N_{n,s}\right\} =\sum_{\left\{ n,s\right\} }N_{n,s}E_{n,s}
\end{equation}%
where we have%
\begin{equation}
\sum_{\left\{ n,s\right\} }N_{n,s}=N.
\end{equation}%
Therefore, the partition function becomes%
\begin{equation}
\mathcal{Z}\left[ N_{n}\right] =\sum_{\left\{ N_{n,s}\right\} }\exp \left[
-\beta \sum_{n=0}^{\infty }\sum_{s=1}^{2}N_{n,s}E_{n,s}\right] .
\end{equation}%
The grand partition function assumes the form%
\begin{equation}
\Xi =\sum_{N=0}^{\infty }\exp \left( \beta \mu N\right) \sum_{\left\{
N_{n,s}\right\} }\exp \left[ -\beta \sum_{n=0}^{\infty
}\sum_{s=1}^{2}N_{n,s}E_{n,s}\right] ,
\end{equation}%
which can be rewritten as
\begin{equation}
\Xi =\prod_{n}\prod_{s}\left\{ \sum_{N_{n,s}=0}^{1}\exp \left[ -\beta
N_{n,s}\left( E_{n,s}-\mu \right) \right] \right\} .
\end{equation}%
Performing the sum over the two possible occupation numbers, we get%
\begin{equation}
\Xi =\prod_{n}\prod_{s}\left\{ 1+\exp \left[ -\beta\left(
E_{n,s}-\mu \right) \right] \right\} .
\end{equation}
The connection with thermodynamics is made using the grand potential given by%
\begin{equation}
\Phi =-\frac{1}{\beta }\ln \Xi .
\end{equation}%
Replacing $\Xi$, in above expression, we get%
\begin{equation}
\Phi =-\frac{1}{\beta }\sum_{n=0}^{\infty }\sum_{s=1}^{2}\ln \left\{ 1+\exp %
\left[ -\beta \left( E_{n,s}-\mu \right) \right] \right\} .
\label{eq:Gand-potential}
\end{equation}%
Despite we could apply the \textit{Euler--Maclaurin} formula, the results are lengthy and not easy to comprehend their respective phenomenology. Therefore, for the sake of en-lighting our analysis, we provide numerical calculations. Also, we utilize the grand potential to calculate the thermodynamic quantities of interest: mean particle number, internal energy, entropy and heat
capacity:%
\begin{equation}
\mathcal{N}=-\frac{\partial \Phi }{\partial \mu },\phantom{a}\mathcal{U}%
=-T^{2}\frac{\partial }{\partial T}\left( \frac{\Phi }{T}\right) ,\phantom{a}%
S=-\frac{\partial \Phi }{\partial T},\phantom{a}C=T\frac{\partial S}{%
\partial T}.
\end{equation}

In order to calculate all thermodynamics quantities using both methods, we need to perform the sum present in Eqs. $\left( \ref%
{eq:Single-Partition-function}\right)$ and $\left( \ref%
{eq:Gand-potential}\right)$. Unfortunately, it is not possible to do so in a closed form to either cases. However, numerical analyses can be performed instead
and we can acquire an idea of the behavior of all thermodynamic quantities, i.e., both in high and low temperature regimes. The next sections are devoted to
study both configurations, namely $d_{ij}$ and $d_{00}$, for a numerical viewpoint. For this propose, we use $r_{0}=50~\mathrm{nm}$ for the radius of the quantum ring \cite{EXP1,EXP2}, $\mathrm{m}=0.511~\mathrm{MeV}$ for the electron mass and $\mathrm{\mu}=0.1~\mathrm{eV}$ for the chemical potential.

\section{Results for Canonical Approach}

This section aims at presenting some numerical result concerning the canonical ensemble formalism. We intend to study the main thermodynamic functions, namely, the Helmholtz free energy, internal energy, entropy and, finally, the heat capacity. In the following subsections, we present the results for two configurations, i.e., the $d_{ij}$ and $d_{00}$. We finish this section showing how the $z$-component of the spin current behave for both configurations as a function of temperature.

\subsection{Configuration $d_{ij}$}

\begin{figure}[tbh]
  \centering
  \subfloat[Helmholtz free energy]{\includegraphics[width=8cm,height=5cm]{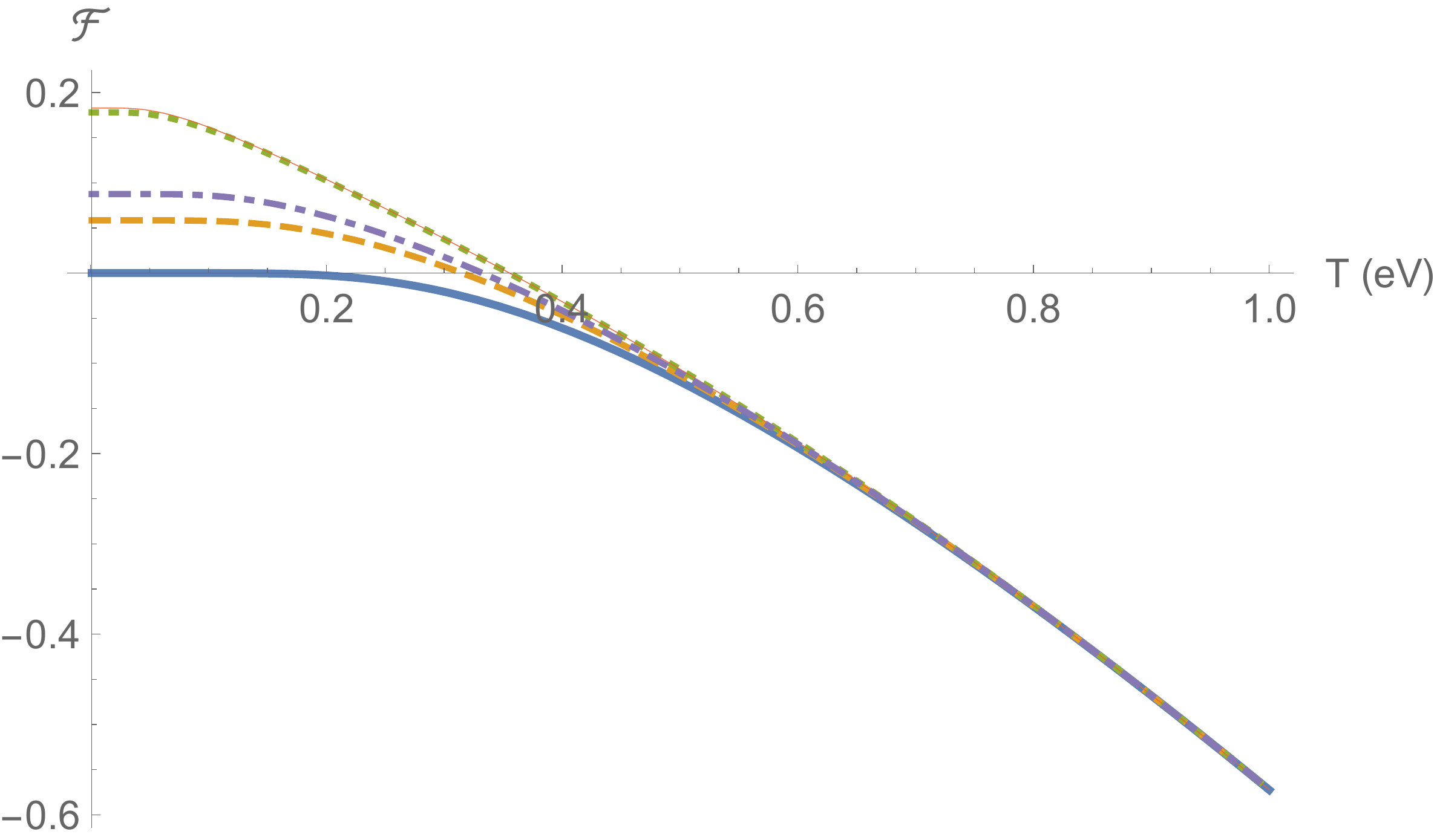}
  \label{fig:Case1-lowTF}}
  \subfloat[Internal energy]{\includegraphics[width=8cm,height=5cm]{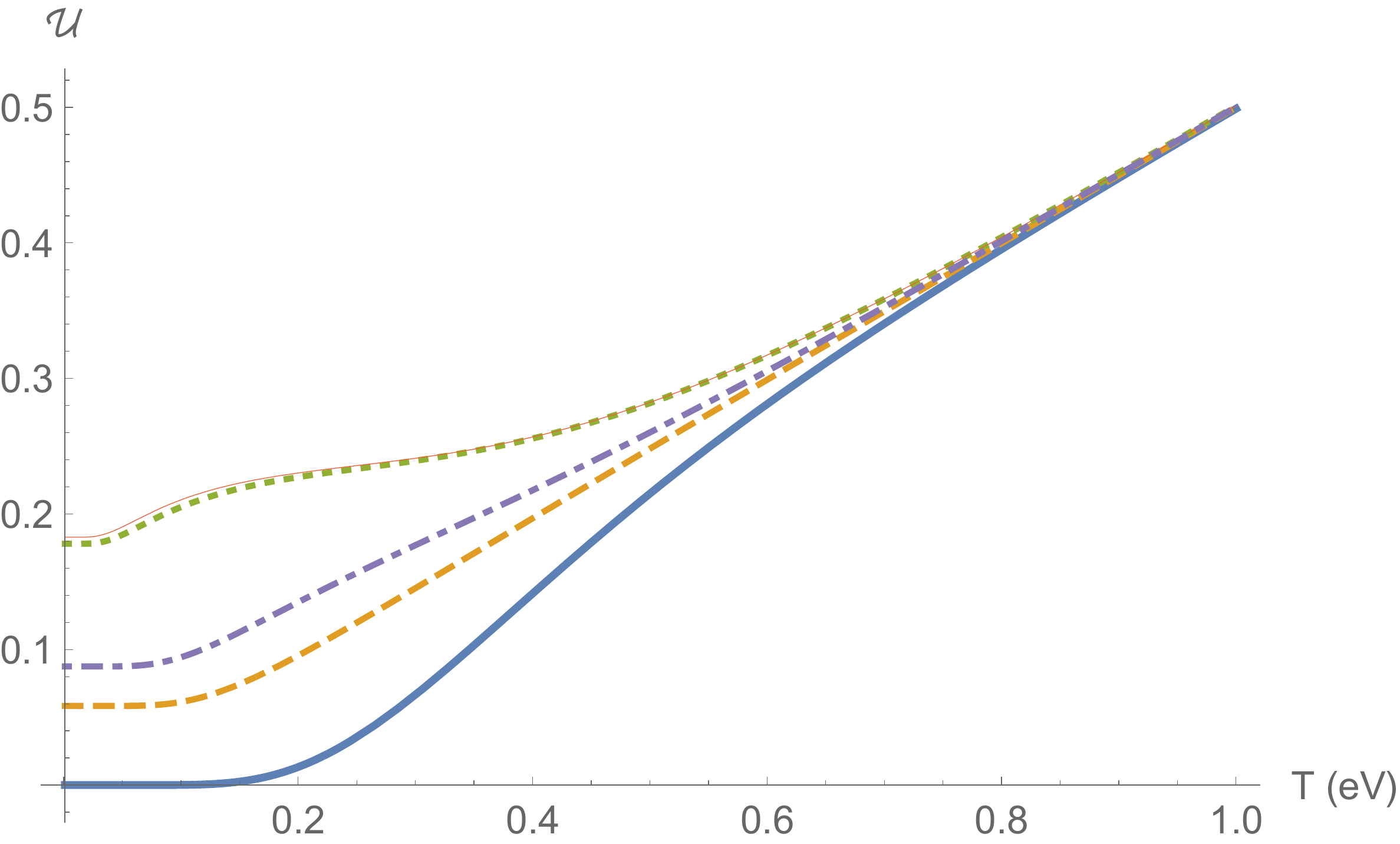}
  \label{fig:Case1-lowTU}}\\
  \subfloat[Entropy]{\includegraphics[width=8cm,height=5cm]{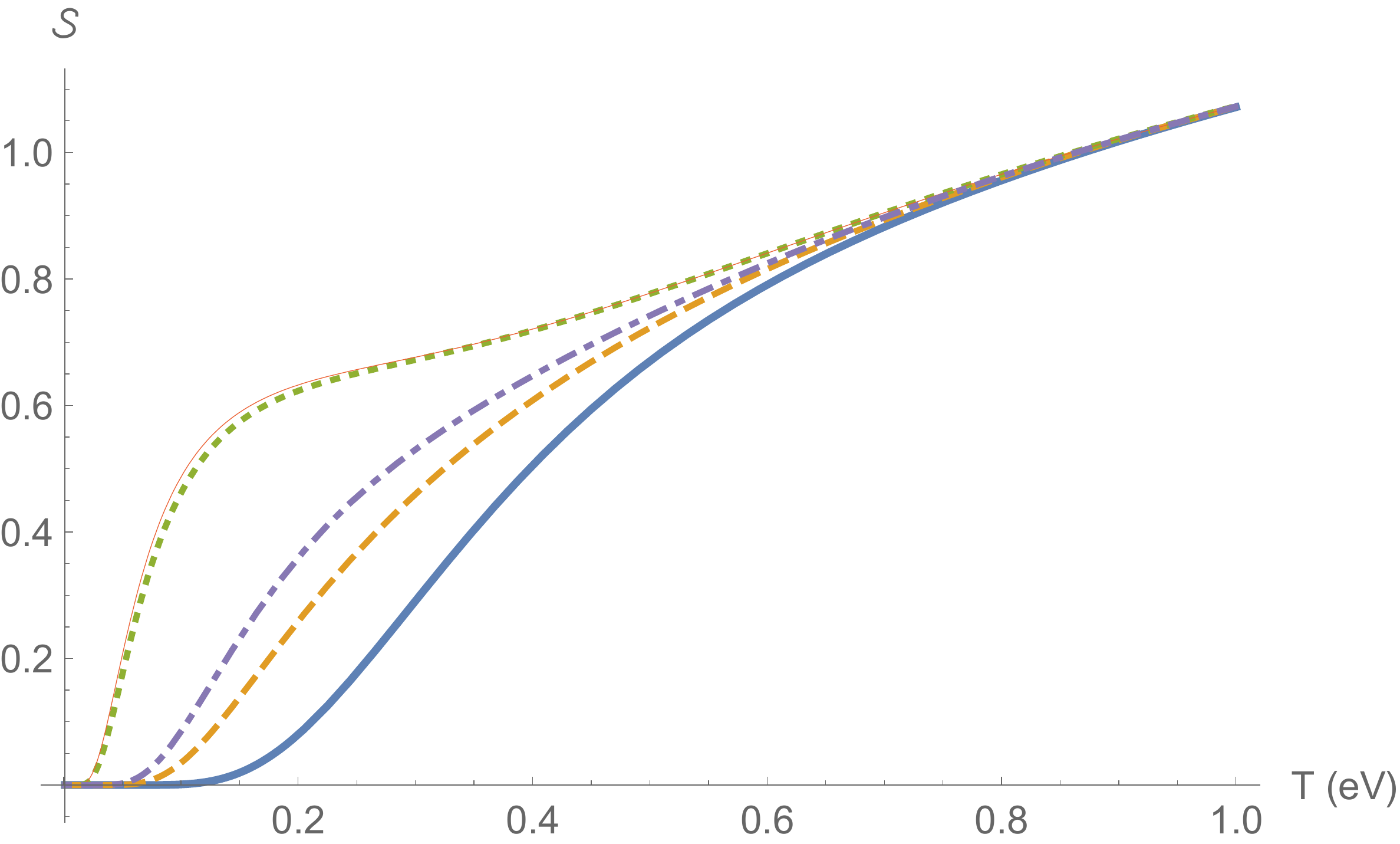}
  \label{fig:Case1-lowTS}}
  \subfloat[Heat capacity]{\includegraphics[width=8cm,height=5cm]{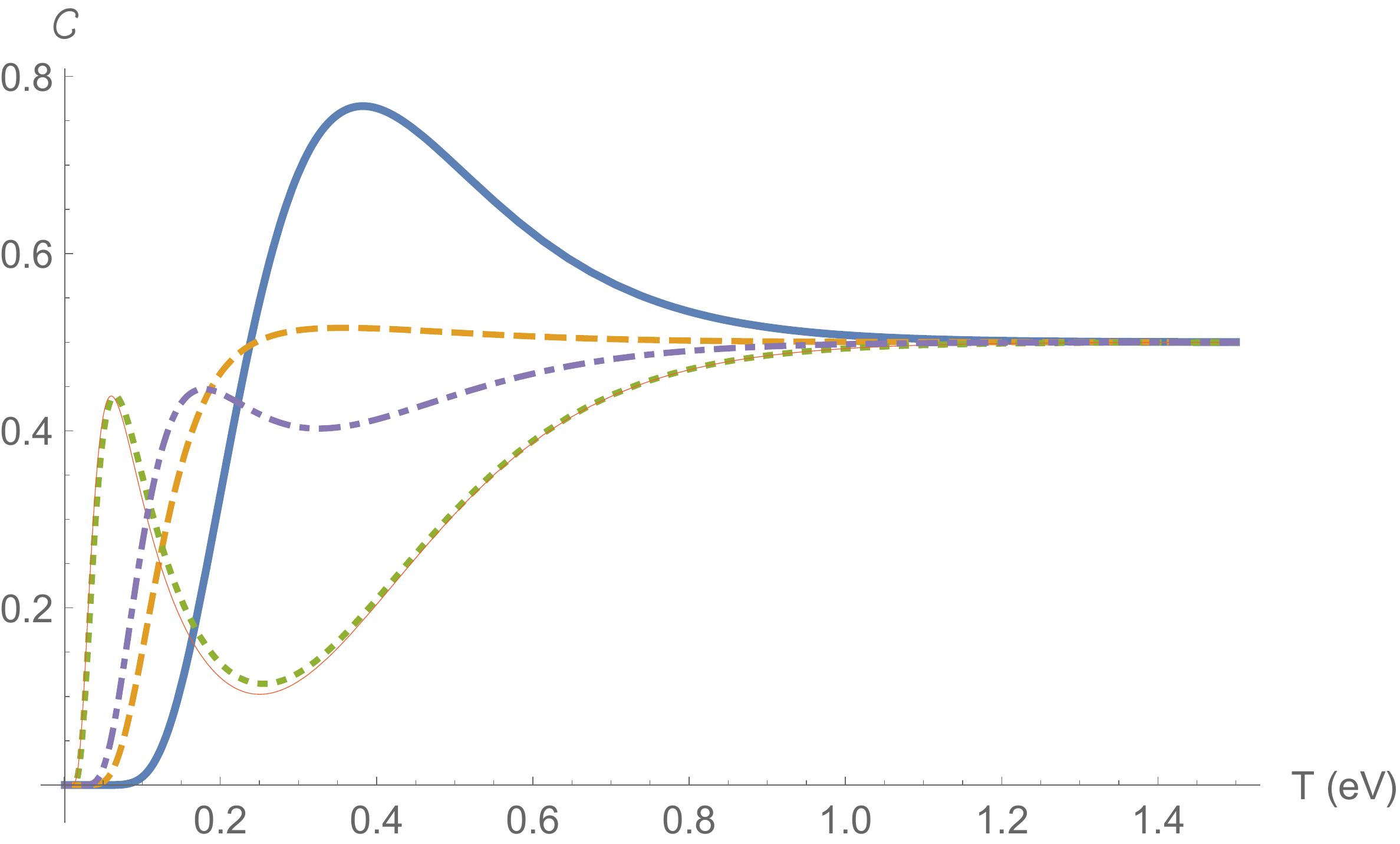}
  \label{fig:Case1-lowTC}}
  \caption{The main thermodynamic functions modified by the anisotropic parameter $d_{ij}$ are displayed above. The plots can be distinguished by the following labels: (thick, blue, $\xi=0.0$), (dashed, orange, $\xi=0.3$), (dotted, green, $\xi=0.6$), (thin, red, $\xi=0.9$), (dotdashed, purple, $\xi=1.2$). More so, the crossover to negative energies happens in larger temperatures for increasing $\xi \in [0,0.9]$.}\label{Case1-lowT}
\end{figure}

The plots displayed in Fig. \ref{Case1-lowT} show the main thermodynamic quantities as a function of temperature. We are also free to choose values for the parameter $\xi$ in order to comprehend how those quantities change. It is important to point out that we consider the low temperature regime the range $0<T<0.6~\mathrm{eV}$ and the high temperatures $T>0.6~\mathrm{eV}$.

Our investigation starts with the Helmholtz free energy shown in Fig. \ref{fig:Case1-lowTF}. In this case, for temperatures relying on the range $0<T<0.6~\mathrm{eV}$, we see that the free energy has its minimum when $\xi=0$, and its maximum when $\xi=0.9$. For $T>0.6~\mathrm{eV}$ all curves tend to the same values. This feature is expected since, for high temperatures, Lorentz violating parameters turn out to be negligible. As the temperature increases, the Helmholtz free energy reaches negative values.

The internal energy is displayed in Fig. \ref{fig:Case1-lowTU}. As before, for $\xi=0.9$, we find the maximum value for the energy. For high temperatures, we obtain the plots in a such way that they converge to the same values. For low temperature regime, we see the most significant difference among the graphics. We observe that $\xi=0.9$ give us the highest energy and, in the range $0<\xi<0.9$, the energy increases as the parameter $\xi$ increases. It is interesting to notice that from $\xi>1.0$ the internal energy decreases and the numerical analyses show us that it keeps decreasing as parameter $\xi$ increases. The entropy showed in Fig. \ref{fig:Case1-lowTS} has the same behavior, i.e., it has its maximum values bounded from $\xi=0.9$, and for high temperature, Lorentz parameter plays no role.

We finish this subsection discussing the behavior of the heat capacity. As displayed in Fig. \ref{fig:Case1-lowTC}, we observe that, for high temperatures, the parameter $\xi$ plays no role in the value of the heat capacity since all plots converge to $\mathcal{C} = 0.5$. On the other hand, for low temperatures, we verify an interesting feature. For instance, for parameters $\xi=0.6$ and $\xi=0.9$, we have a local minimum at $T=0.25~\mathrm{eV}$. Moreover, for $\xi=0$, we have a local maximum at $T=0.4~\mathrm{eV}$, and for $\xi=0.3$, the local maximum behavior fades. For the ranges $0<\xi<0.9$ and $0.2~\mathrm{eV}<T<0.4~\mathrm{eV}$, we observe a transitions between a local maximum to a local minimum. For $\xi>0.9$, we see the formation of a new local maximum values in the range $0.15~\mathrm{eV}<T<0.2~\mathrm{eV}$. For temperature below $0.15~\mathrm{eV}$, we see that both $\xi=0.6$ and $\xi=0.9$ have a local maximum at $0.05~\mathrm{eV}$, while for the other values of the parameter $\xi$, the heat capacity goes to zero.

\subsection{Configuration $d_{00}$}

\begin{figure}[tbh]
  \centering
  \subfloat[Helmholtz free energy]{\includegraphics[width=8cm,height=5cm]{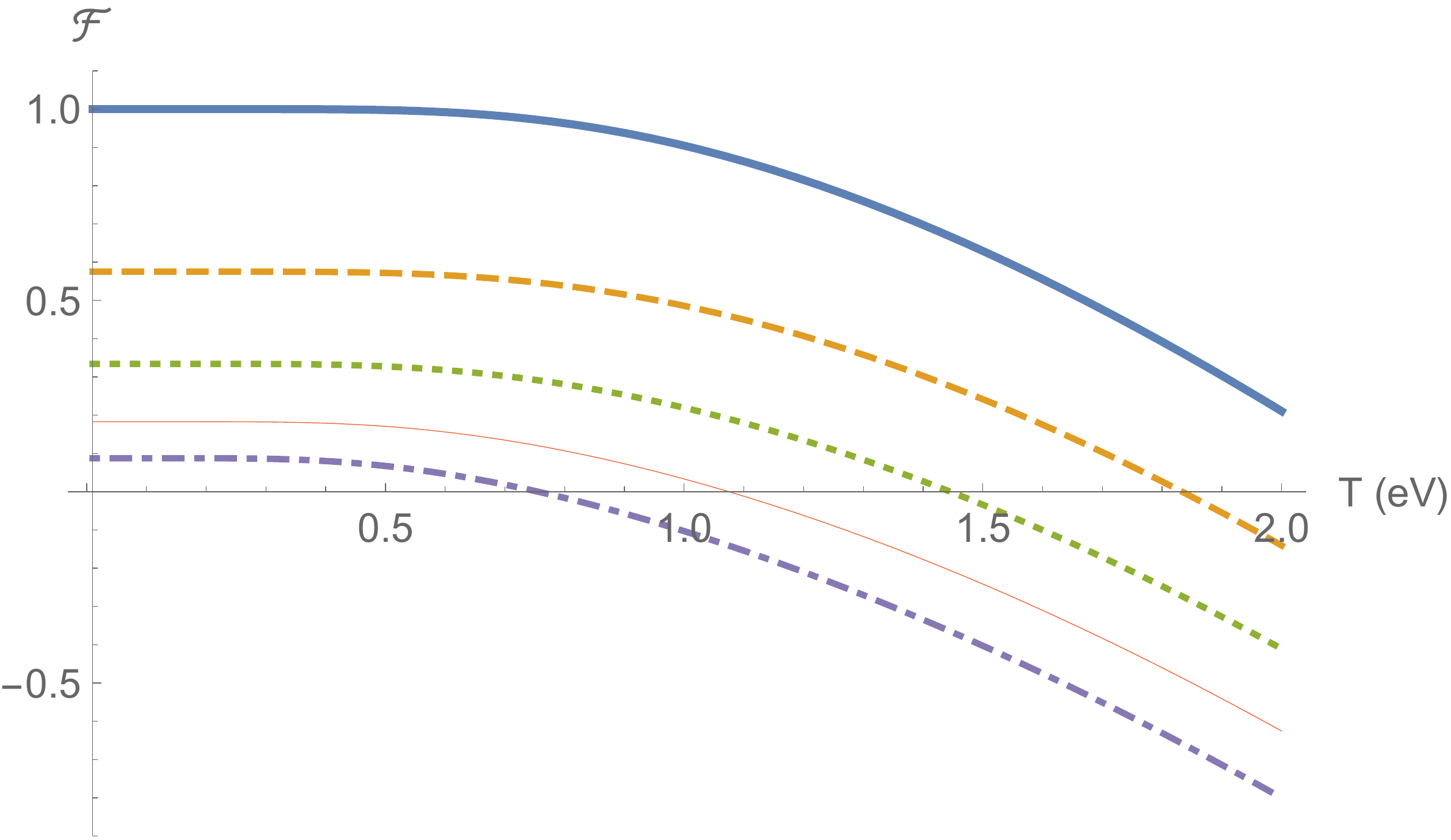}
  \label{fig:Case2-lowTF}}
  \subfloat[Internal energy]{\includegraphics[width=8cm,height=5cm]{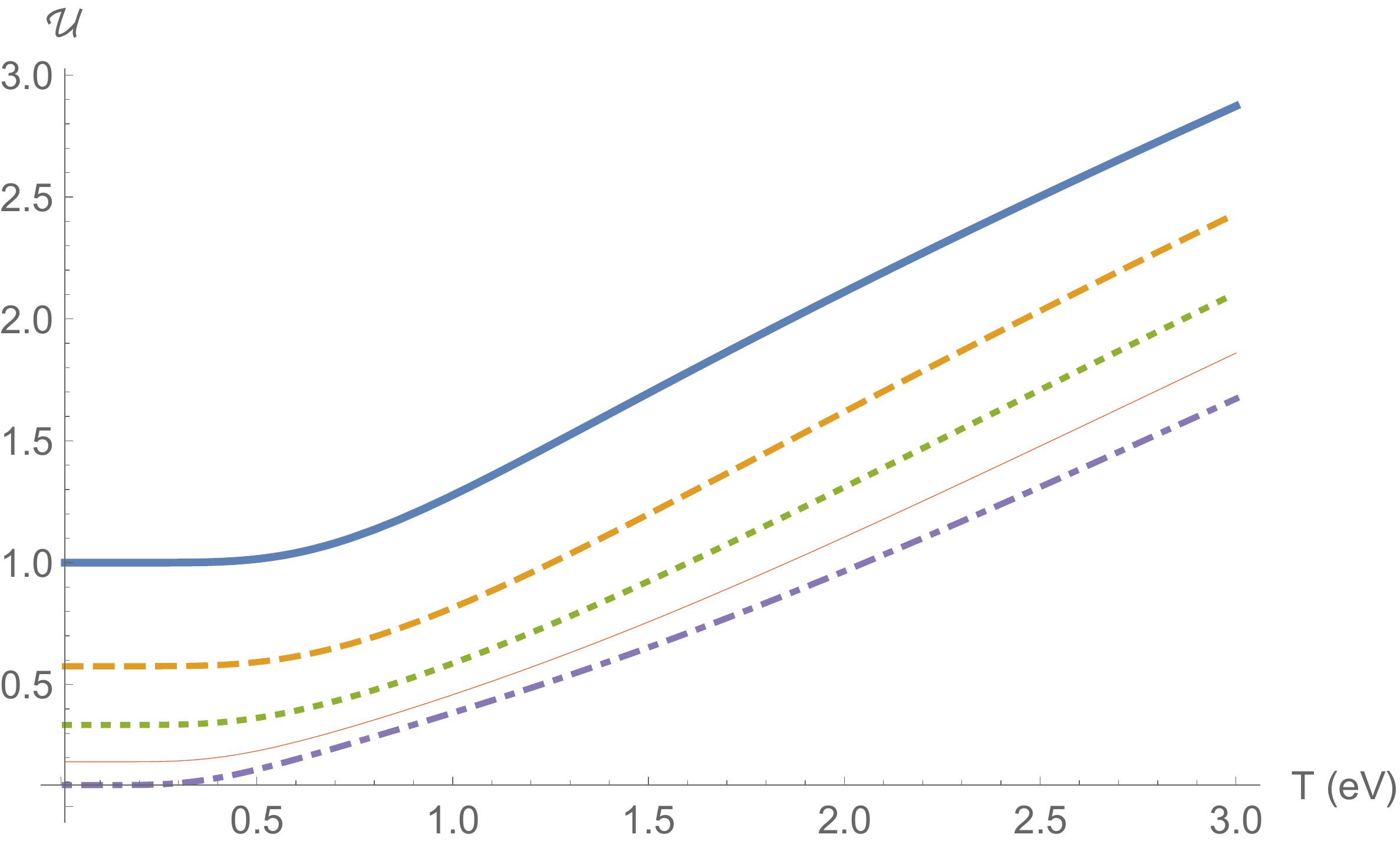}
  \label{fig:Case2-lowTU}}\\
  \subfloat[Entropy]{\includegraphics[width=8cm,height=5cm]{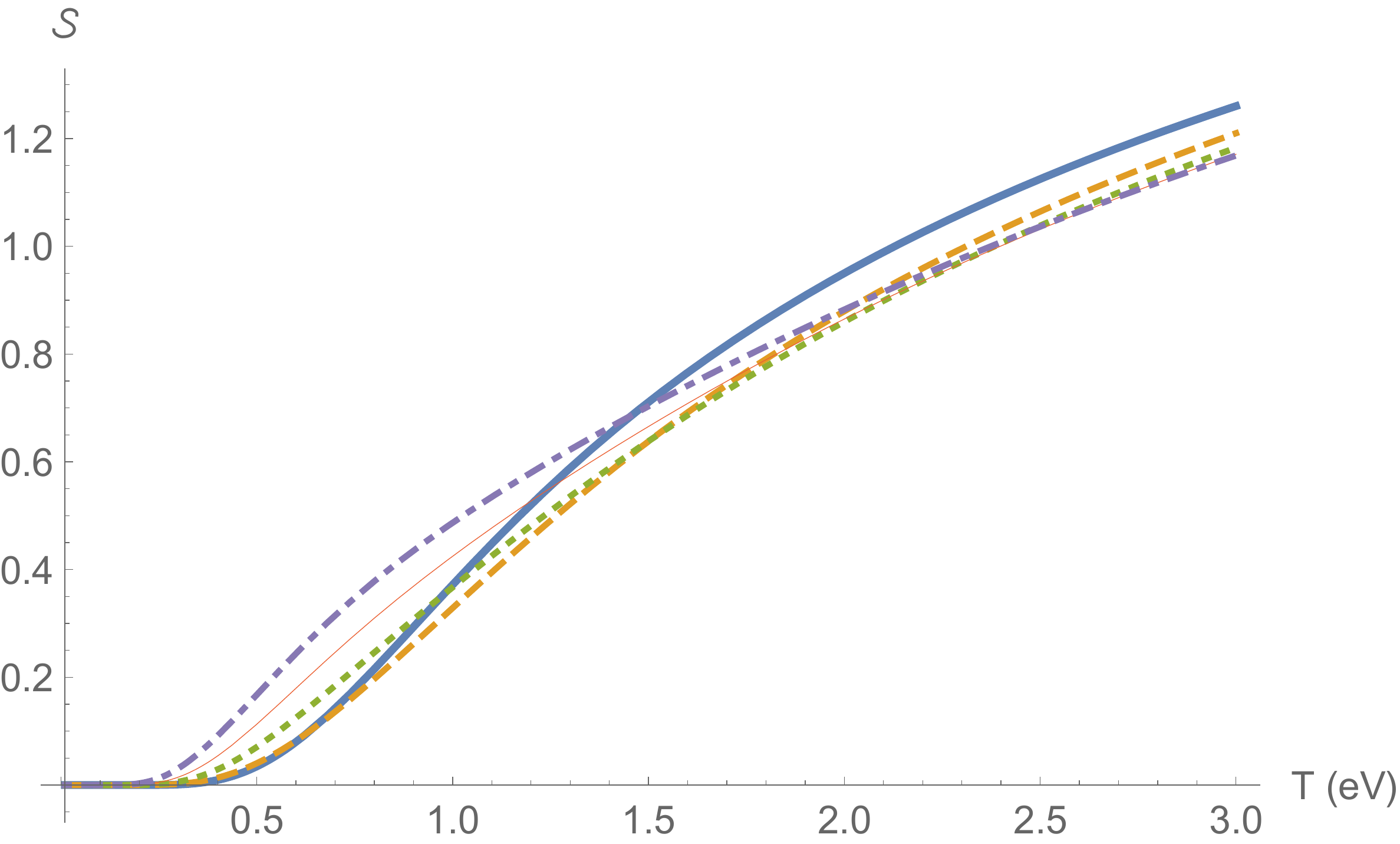}
  \label{fig:Case2-lowTS}}
  \subfloat[Heat capacity]{\includegraphics[width=8cm,height=5cm]{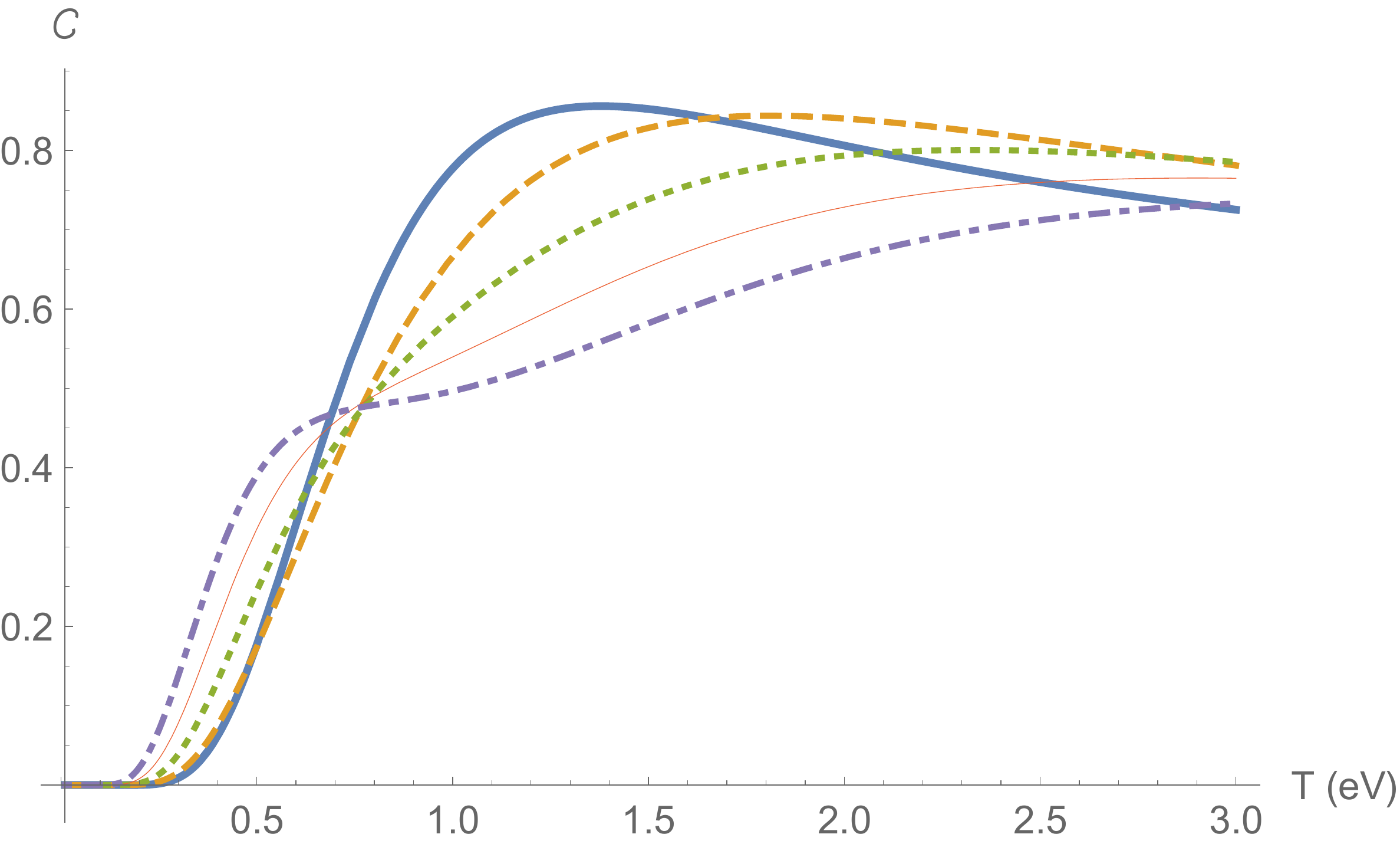}
  \label{fig:Case2-lowTC}}
  \caption{The thermodynamic behavior of Helmholtz free energy, internal energy, entropy and heat capacity, all modified by the isotropic parameter $d_{00}$, are presented. The plots can be identified using the labels as follows: (thick, blue, $\xi_{00}=0.0$), (dashed, orange, $\xi_{00}=0.3$), (dotted, green, $\xi_{00}=0.6$), (thin, red, $\xi_{00}=0.9$), (dotdashed, purple, $\xi_{00}=1.2$).}\label{fig:Case2-lowT}
\end{figure}

In this subsection, we intend to study the modifications produced by the Lorentz violating parameter $d_{00}$. The plots displayed in Fig. \ref{fig:Case2-lowT} summarize the main thermodynamic quantities as a function of temperature. We also use the freedom of the parameter $\xi_{00}$ in order to understand how the thermodynamic quantities change with both temperature and $\xi_{00}$. 

Now, let us start with the Helmholtz free energy shown in Fig. \ref{fig:Case2-lowTF}. The first feature is that the Helmholtz free energy decreases as parameter $\xi_{00}$ increases. In this way, we conclude that the free energy is bounded from above for $\xi_{00}=0$. We notice that as parameter $\xi_{00}$ increases, the difference between energies, i.e, $\mathcal{F}(\xi_{00})-\mathcal{F}(\xi_{00}+0.3)$, decreases. We also see that the influence of the parameter $\xi_{00}$ persists for temperatures greater than those ones presented in Ref. \ref{fig:Case1-lowTF}. Nevertheless, this influence vanishes whether we reach high temperatures. A similar behavior is verified for the internal energy, Fig. \ref{fig:Case2-lowTU}. We detect that energy is a convex function of the temperature.

In particular, the behavior of entropy is exhibited in Fig. \ref{fig:Case2-lowTS}. We note that such thermodynamic function is a concave function of temperature. Differently of what happened in the previous case, there is no value of parameter $\xi_{00}$ that bounds the entropy either from above or below. The entropy tends to zero while the temperature goes to zero. For high temperature regime, the parameter $\xi_{00}$ plays no role in such a case. In other words, the entropy assumes the same value no matter if $\xi_{00}$ increases or decreases. We also observe that the entropy increases for $\xi_{00}>0$.

Finally, we focus on the heat capacity encountered in Fig. \ref{fig:Case2-lowTC}. For the high temperature regime, the heat capacity goes to $0.7$, and, in the range $0<T<0.7~\mathrm{eV}$, this thermodynamic function is a crescent function of the parameter $\xi_{00}$. However, this behavior changes if $T\approx0.7~\mathrm{eV}$. For $T>0.7~\mathrm{eV}$, the heat capacity is a decreasing function of $\xi_{00}$. It is important to notice that, at $\xi_{00}=0$, the heat capacity has a maximum. We also see the convexity flattens as the parameter increases.

\subsection{Spin currents}

As we have already mentioned, we can also study the behavior of the spin current $\mathcal{J}_{\varphi }^{z}$ as a function temperature. To do that, we need to consider the fluctuations of the canonical ensemble:
\begin{align}
\left\langle \mathcal{J}_{\varphi }^{z}\right\rangle & =\frac{1}{\mathcal{Z}%
_{1}}\sum_{n=0}^{\infty }\sum_{\left\{ s,\lambda \right\} =\left\{ \pm
\right\} }\mathcal{J}_{\varphi }^{z}\exp \left( -\beta E_{n,s}^{\lambda
}\right) ,\displaybreak[0] \notag\\
& =\frac{1}{\mathcal{Z}_{1}}\sum_{n=0}^{\infty }\sum_{\left\{ s,\lambda
\right\} =\left\{ \pm \right\} }\frac{1}{4\mathrm{m}r_{0}}\left( 2n\cos \theta
-1\right) \exp \left( -\beta E_{n,s}^{\lambda }\right) ,\displaybreak[0]\notag \\
& =\frac{2\cos \theta }{4\mathrm{m}r_{0}}\frac{1}{\mathcal{Z}_{1}}\sum_{n=0}^{\infty
}\sum_{\left\{ s,\lambda \right\} =\left\{ \pm \right\} }n\exp \left( -\beta
E_{n,s}^{\lambda }\right) -\frac{1}{4\mathrm{m}r_{0}}.
\end{align}%

\begin{figure}[tbh]
  \centering
  \subfloat[Spin current for $d_{ij}$ ]{\includegraphics[width=8cm,height=5cm]{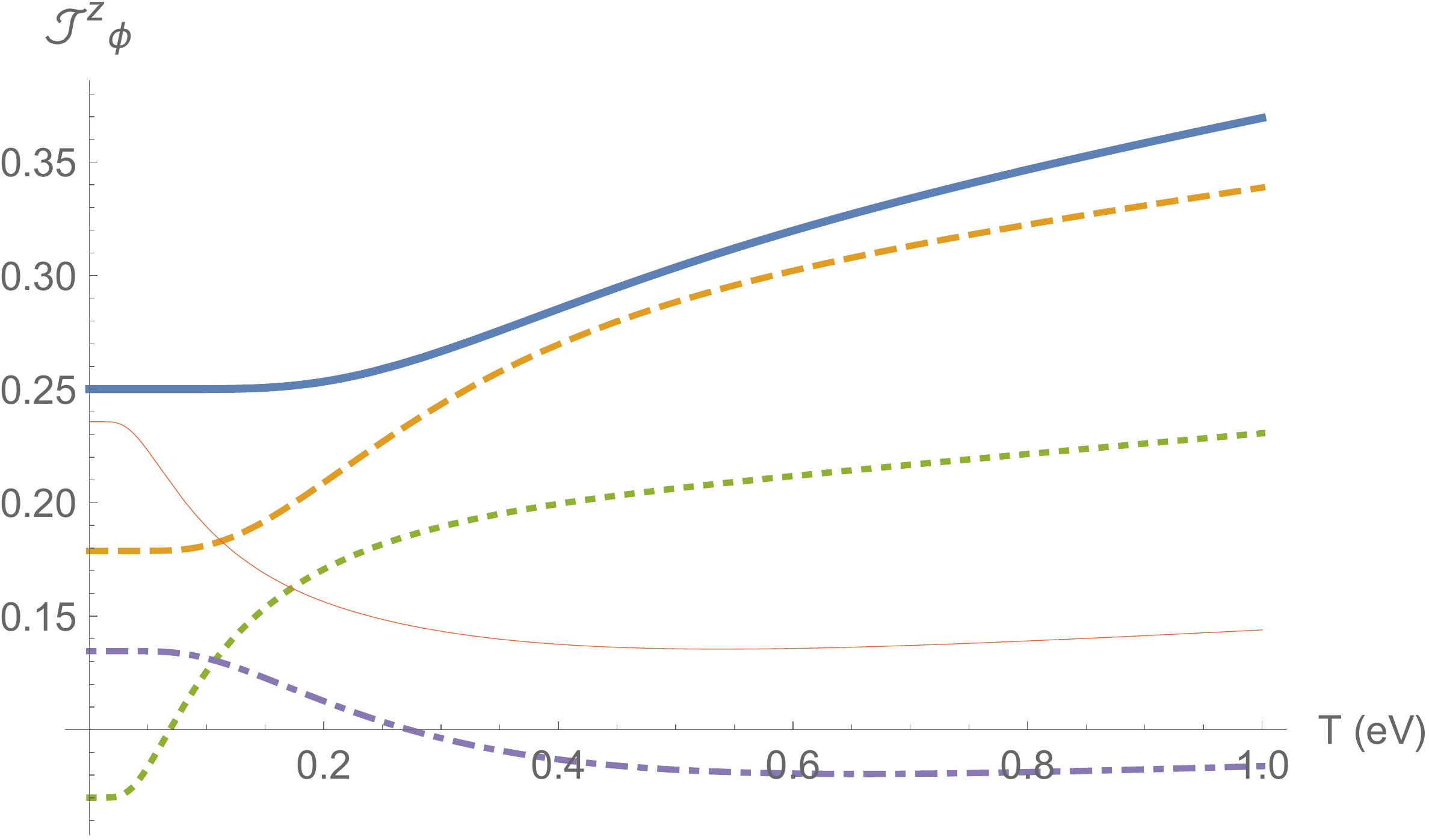}
  \label{fig:CCase1-f1}}
  \subfloat[Spin current for $d_{00}$ ]{\includegraphics[width=8cm,height=5cm]{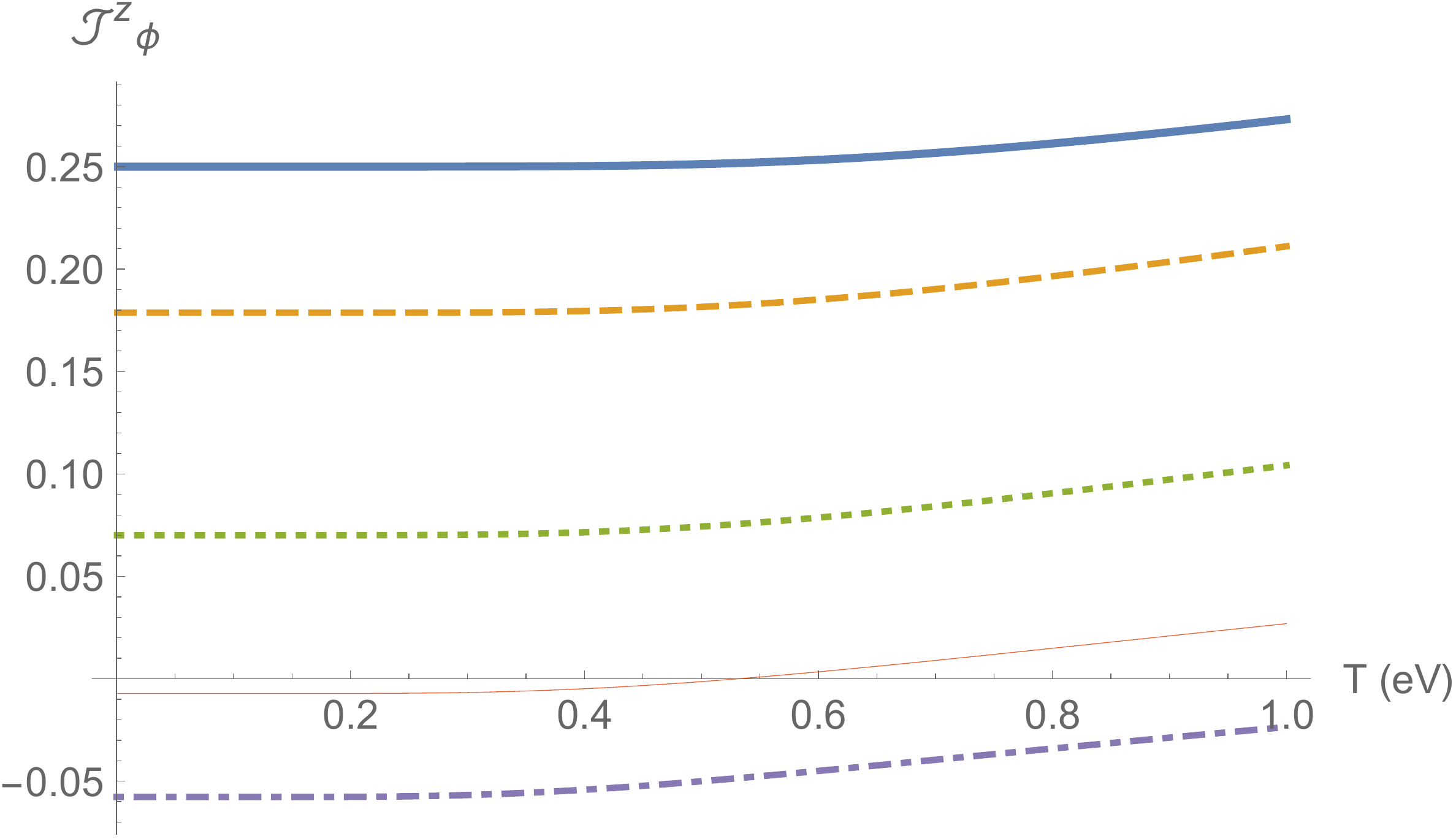}
  \label{fig:CCase1-f2}}
  \caption{The spin currents results for the canonical approach are displayed above, where the labels (thick, blue, $\{\xi,\xi_{00}\}=0.0$), (dashed, orange, $\{\xi,\xi_{00}\}=0.3$), (dotted, green, $\{\xi,\xi_{00}\}=0.6$), (thin, red, $\{\xi,\xi_{00}\}=0.9$), (dotdashed, purple, $\{\xi,\xi_{00}\}=1.2$) help us to identify the values chosed for the parameter $\xi$.}\label{fig:CCaseJ}
\end{figure}

In Fig. \ref{fig:CCaseJ}, it is shown how $z$-component of spin currents behave for some values of temperature. Initially, let us analyze $d_{ij}$-case exhibited in Fig. \ref{fig:CCase1-f1}. Here, when $T>0.2~\mathrm{eV}$, the spin current is a decreasing function of $\xi$. Nevertheless, when $T<0.2~\mathrm{eV}$, something unusual emerges.

On the other hand, for the range $0<\xi<0.6$, the spin current has its values diminished when parameter $\xi$ increases. Nevertheless, when it overcomes $\xi = 0.6$, the behavior changes and $\mathcal{J}_{\varphi }^{z}$ increases assuming a value grater than $\xi=0.3$. Therefore, it recovers behavior of a decreasing function of $\xi$, with different range $0.7<\xi<1.2$ though.

Now, let us study the implications of temperature and $\xi$ at the current modified by $d_{00}$, see Fig. \ref{fig:CCase1-f2}. We observe at first that the current $\mathcal{J}_{\varphi }^{z}$ is a function of temperature. We also realise the fact that the current is a decreasing function of $\xi$ and the difference between two values of the current, that is $\mathcal{J}_{\varphi }^{z}(T,\xi_{1})-\mathcal{J}_{\varphi }^{z}(T,\xi_{2})$, is kept independent of the temperature. The difference in the behavior of the two cases displayed in Fig. \ref{fig:CCaseJ} relies on the fact that the parameter $d_{ij}$ induces an anisotropy in the space whereas the parameter $d_{00}$ is isotropic.

\section{Results for Grand canonical Approach}

In this section, we present numerical analyses based on the grand canonical ensemble formalism. Particularly, we show the results for the particle number, internal energy, entropy and the heat capacity as a function of temperature and $\xi$. In what follows, the outcome results considering again $d_{ij}$ and $d_{00}$ are exhibited. Analogously to the previous section, we finish our discussion, showing how $z$-component of the spin current interferes in our system as a function of the temperature.

\subsection{Configuration $d_{ij}$}

\begin{figure}[tbh]
  \centering
  \subfloat[Number of particles]{\includegraphics[width=8cm,height=5cm]{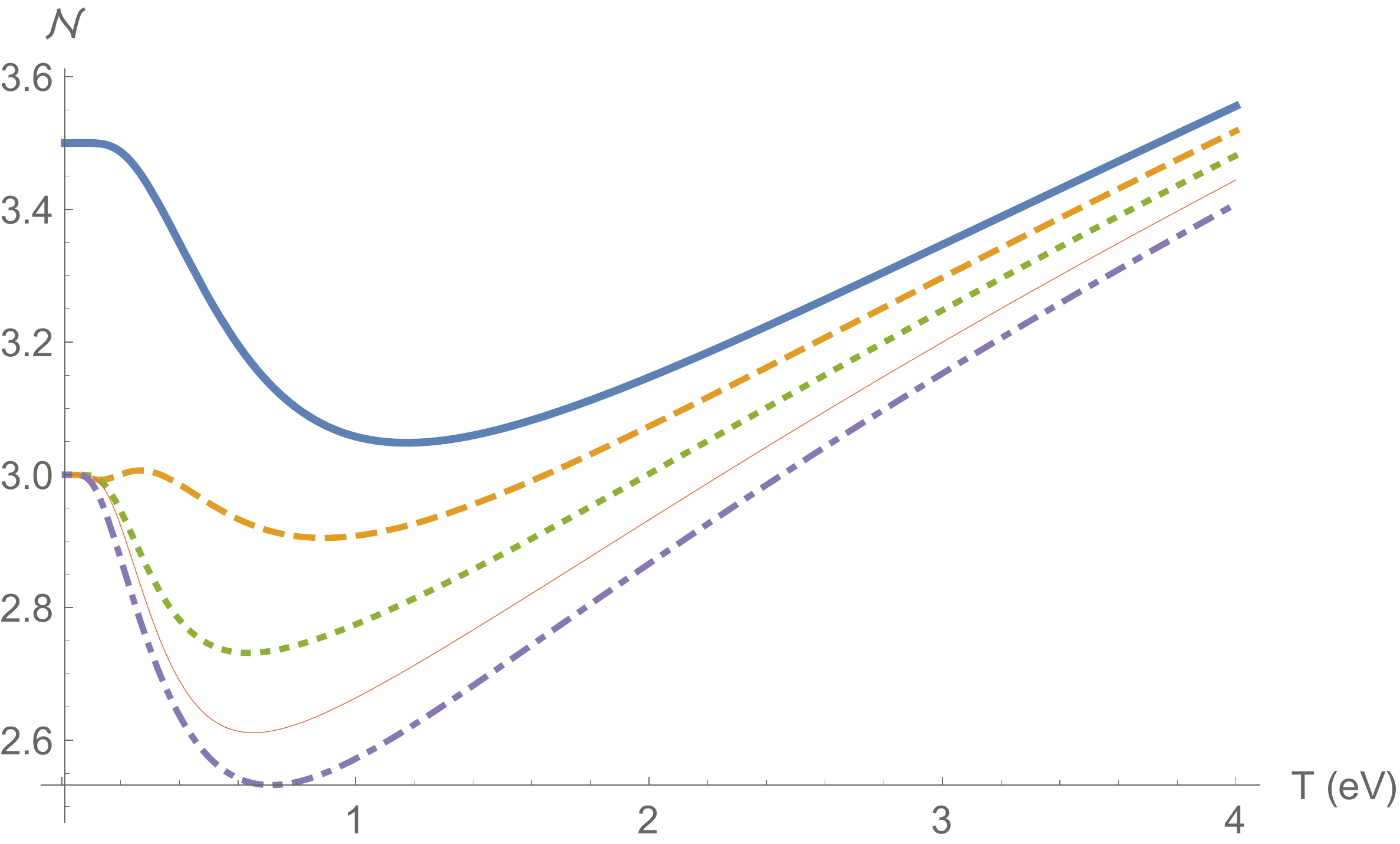}
  \label{fig:GPCase-1-lowTN}}
  \subfloat[Internal energy]{\includegraphics[width=8cm,height=5cm]{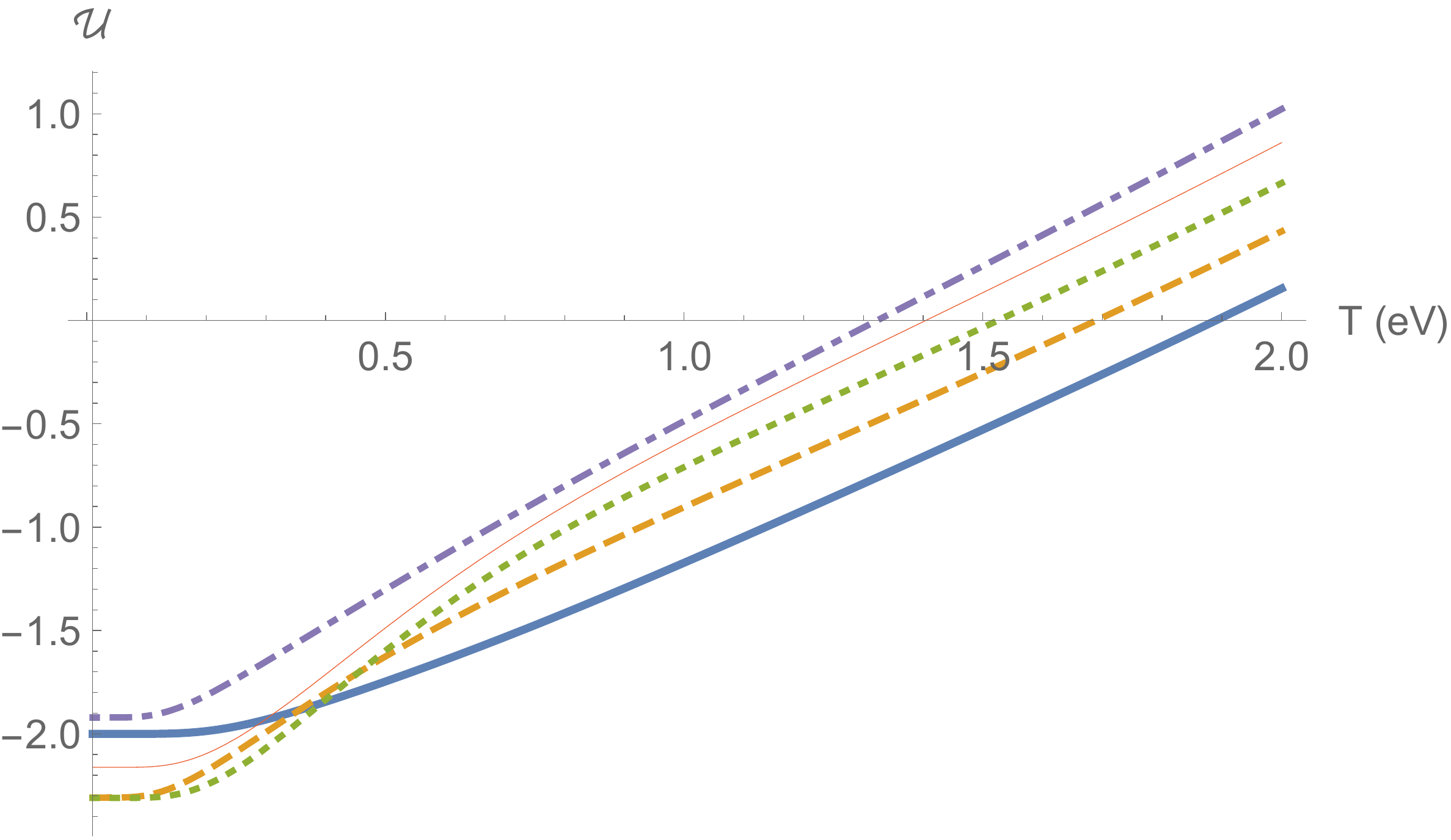}
  \label{fig:GPCase-1-lowTU}}\\
  \subfloat[Entropy]{\includegraphics[width=8cm,height=5cm]{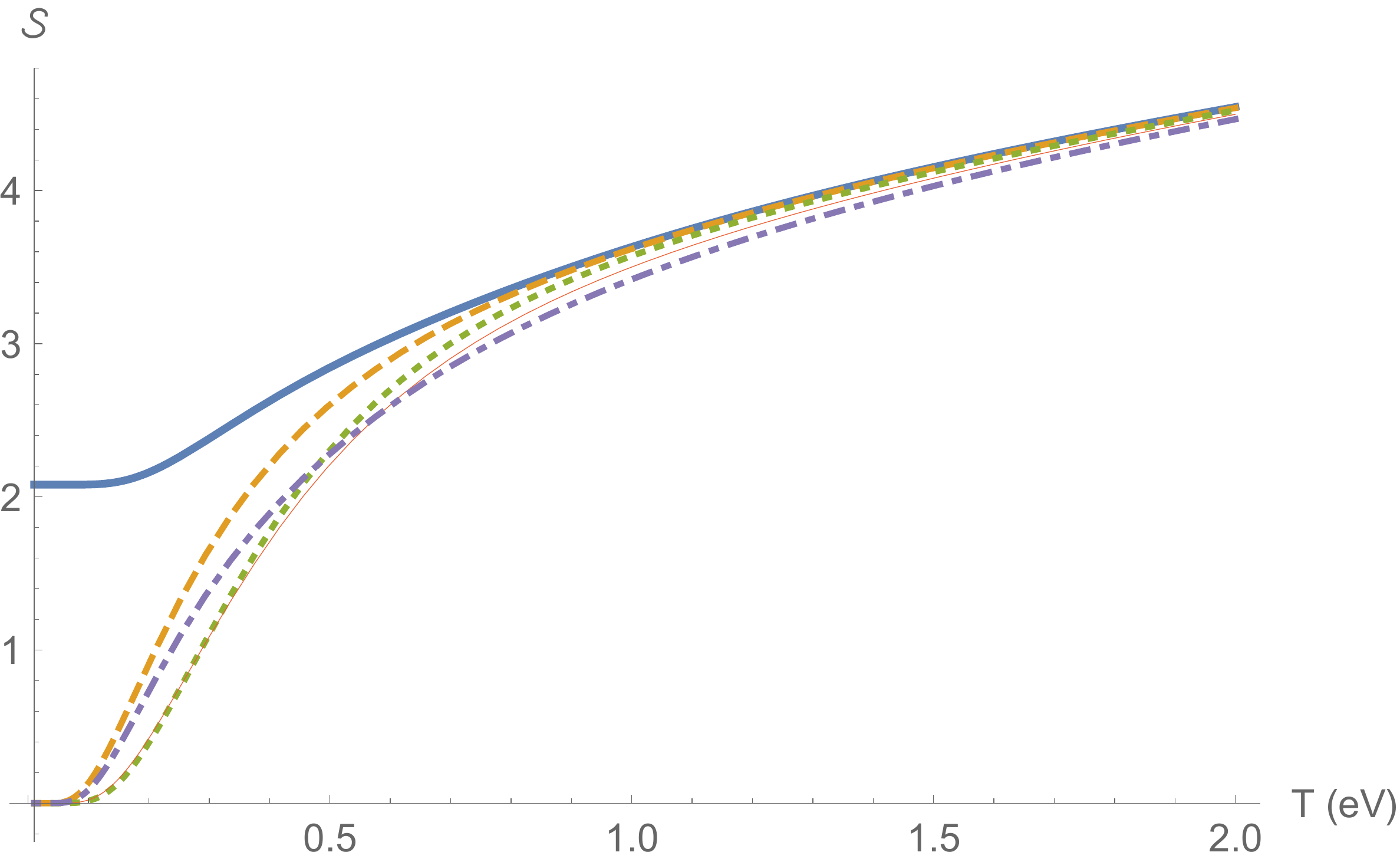}
  \label{fig:GPCase-1-lowTS}}
  \subfloat[Heat capacity]{\includegraphics[width=8cm,height=5cm]{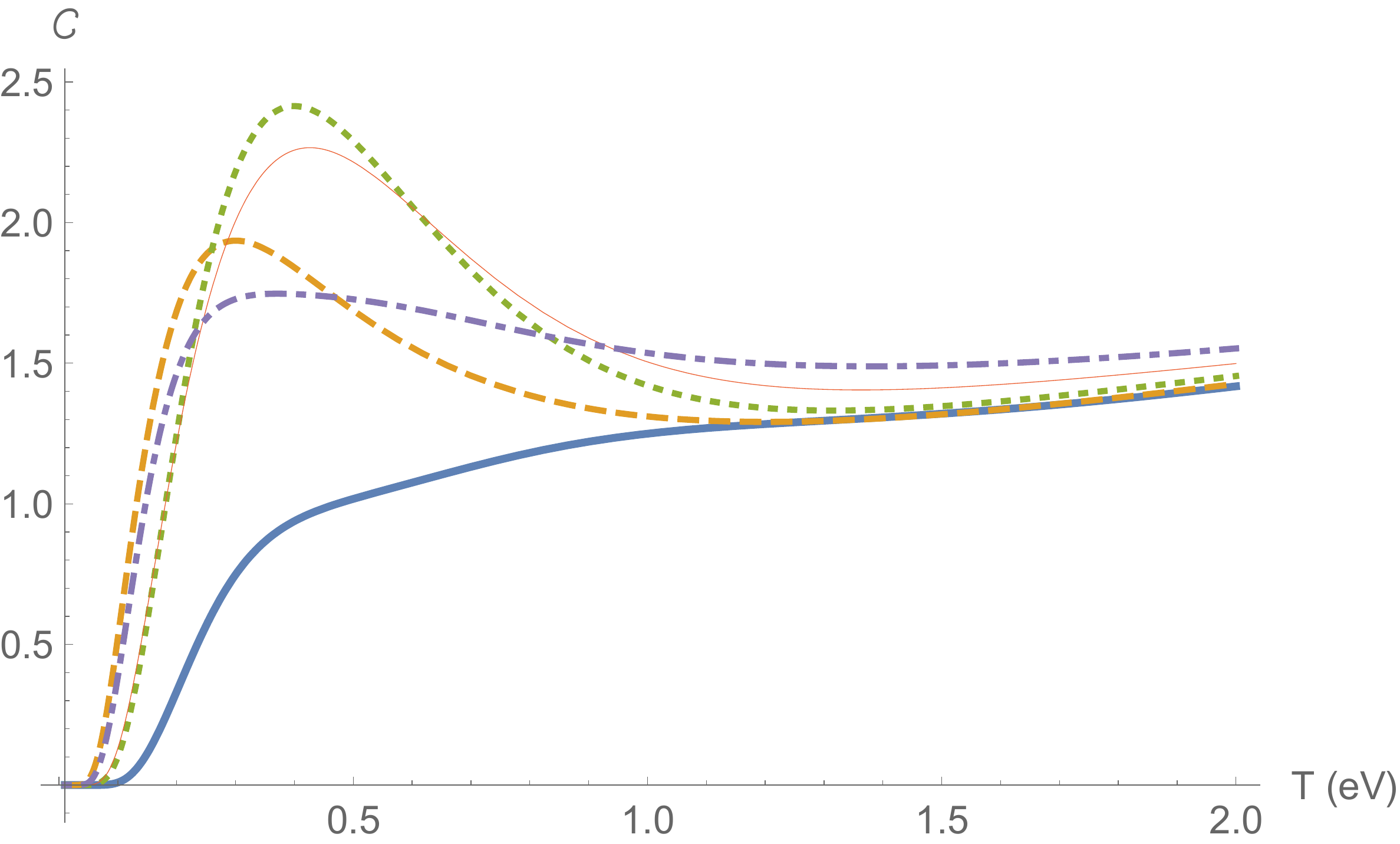}
  \label{fig:GPCase-1-lowTC}}
  \caption{The main thermodynamic functions obtained by the grand canonical approach and modified by the anisotropic parameter $d_{ij}$ are displayed above. The plots can be distinguished by the following labels: (thick, blue, $\xi=0.0$), (dashed, orange, $\xi=0.3$), (dotted, green, $\xi=0.6$), (thin, red, $\xi=0.9$), (dotdashed, purple, $\xi=1.2$).}\label{fig:GPCase-1-LowT}
\end{figure}

Let us start by looking, in Fig. \ref{fig:GPCase-1-lowTN}, how the particle number changes. We observe that $\mathcal{N}(T,\xi)$ is a convex function of temperature, i.e., it has a minimum. Another point is that the function $\mathcal{N}(T,\xi)$ turns out to have no dependence of $\xi$ when high temperatures are taken into account. We also see that such thermal property is a decreasing function of $\xi$ and the minimum is shifted to the left whether $\xi$ increases. Notably, since particles are allowed to vary, we can use such a feat to control the number of electrons inside the ring.

Another important thermodynamic function is the internal energy, see Fig. \ref{fig:GPCase-1-lowTU}. We divide our analyses into two ranges, namely: $T\leq 0.4~\mathrm{eV}$ and $T>0.4~\mathrm{eV}$ for a better comprehension. For the regime of temperature below $T=0.4~\mathrm{eV}$, we verify the following pattern: $\mathcal{U}(T,\xi=0.6)<\mathcal{U}(T,\xi=0.3)<\mathcal{U}(T,\xi=0.9)<\mathcal{U}(T,\xi=0)<\mathcal{U}(T,\xi=1.2)$. On the other hand, for the range $T>0.4~\mathrm{eV}$, we realise that the internal energy is a increasing function of $\xi$.

The entropy is displayed in Fig. \ref{fig:GPCase-1-lowTS}. We see that entropy is a monotonic increasing function of the temperature. For $T=0$, we observe that $\mathcal{S}(T=0,\xi=0)=2$ and for $\mathcal{S}(T=0,\xi\neq0)=0$. When we take into account the Lorentz violating parameter, the entropy has its value shifted to zero independent of $\xi$. We also conclude that $\mathcal{S}(T,\xi)$ tends to be parameter independent for high temperatures.

Finally, let us observe the heat capacity exhibited in Fig. \ref{fig:GPCase-1-lowTC}. We notice that a local maximum is formed when we regard the parameter $\xi$. We also observe that the heat capacity tends to $1.5$ as temperature grows. The maximum value for the heat capacity is obtained for the configuration $\mathcal{C}(T=0.4,\xi=0.6)$.

\subsection{Configuration $d_{00}$}

\begin{figure}[tbh]
\centering
  \subfloat[Number of particles]{\includegraphics[width=8cm,height=5cm]{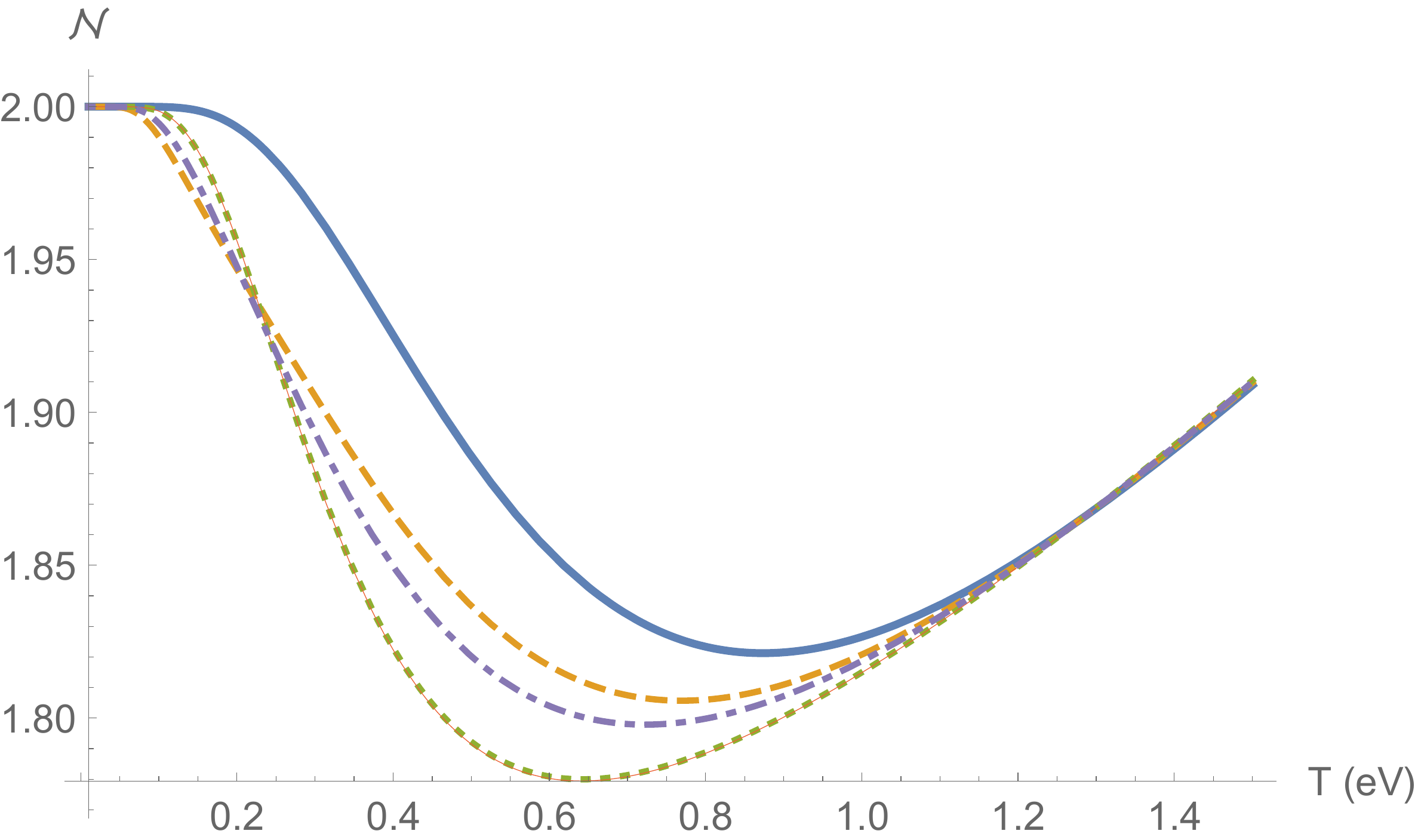}
  \label{fig:GPCase-2-lowTN}}
  \subfloat[Internal energy]{\includegraphics[width=8cm,height=5cm]{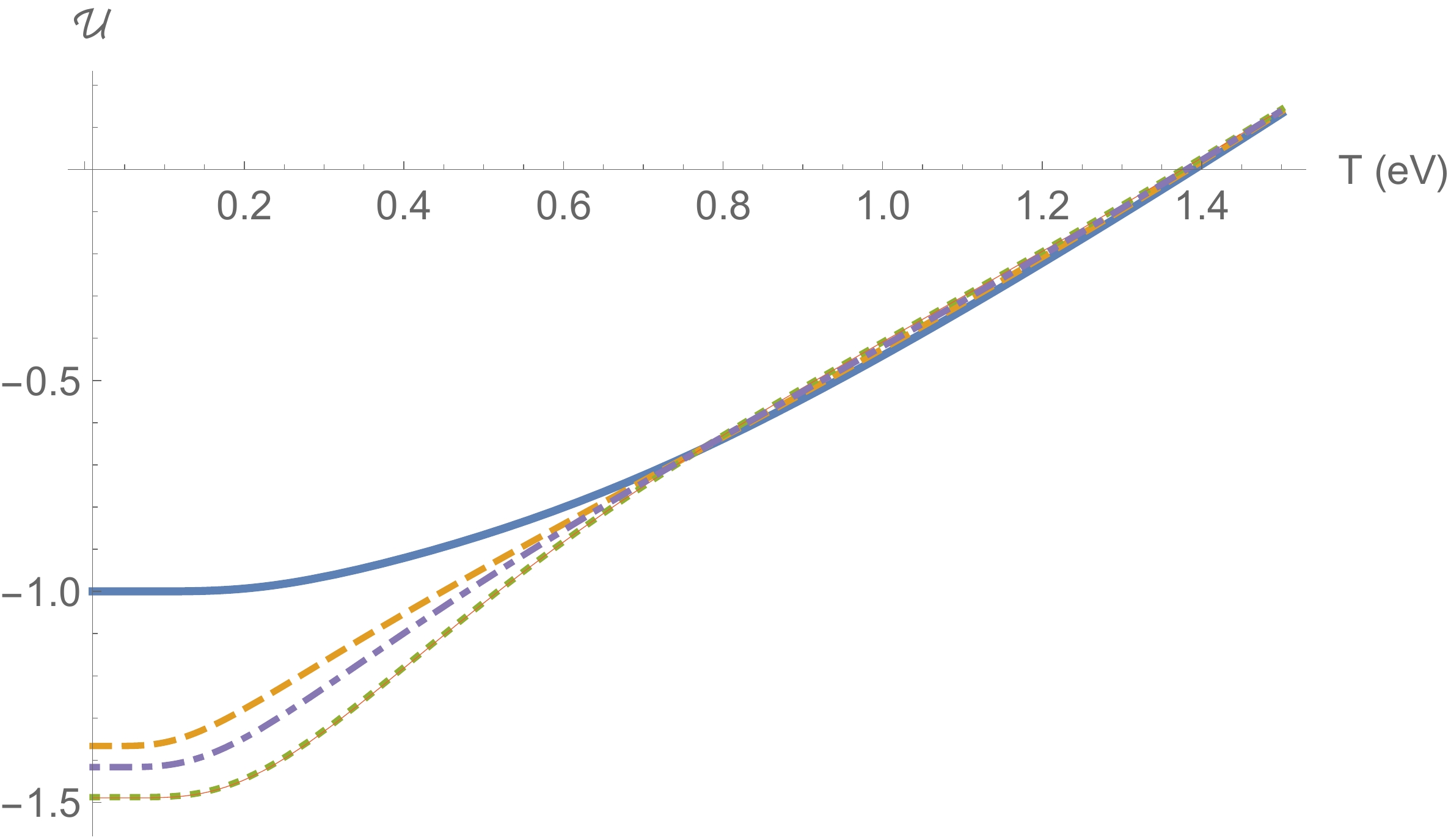}
  \label{fig:GPCase-2-lowTU}}\\
  \subfloat[Entropy]{\includegraphics[width=8cm,height=5cm]{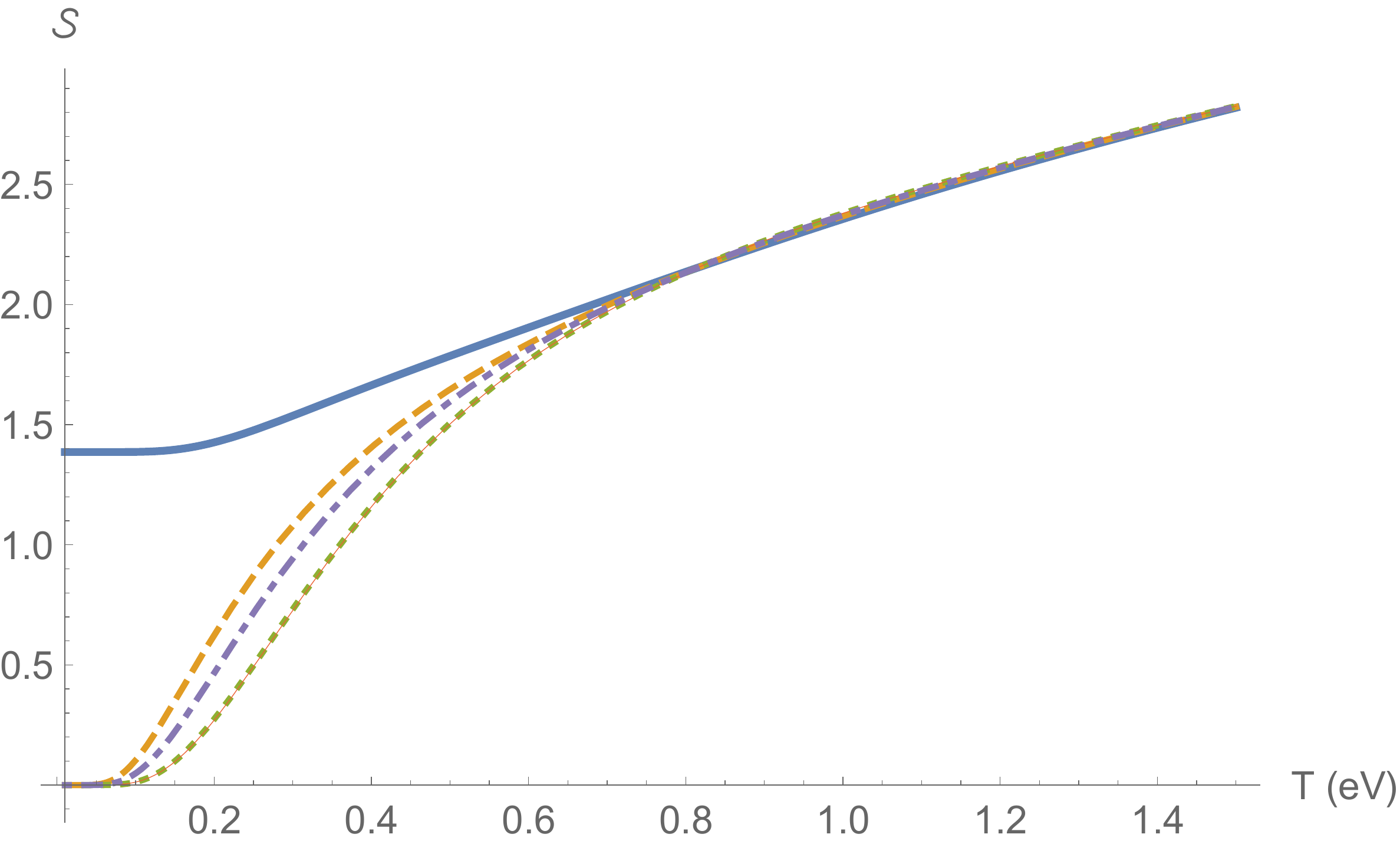}
  \label{fig:GPCase-2-lowTS}}
  \subfloat[Heat capacity]{\includegraphics[width=8cm,height=5cm]{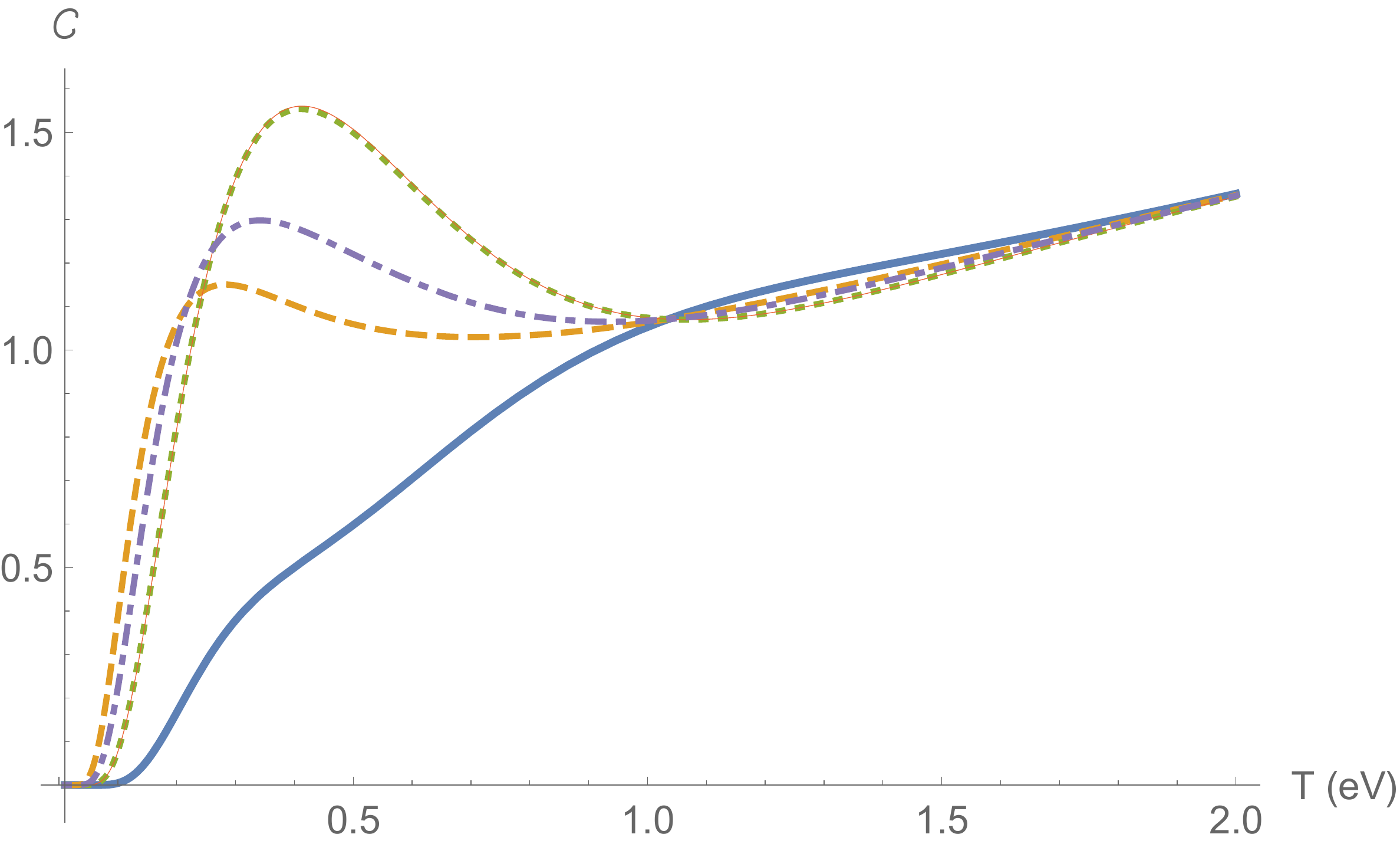}
  \label{fig:GPCase-2-lowTC}}
\caption{The thermodynamic behavior of Helmholtz free energy, internal energy, entropy and heat capacity, all modified by the isotropic parameter $d_{00}$, are showed above. The plots can be identified using the labels: (thick, blue, $\xi_{00}=0.0$), (dashed, orange, $\xi_{00}=0.3$), (dotted, green, $\xi_{00}=0.6$), (thin, red, $\xi_{00}=0.9$), (dotdashed, purple, $\xi_{00}=1.2$).}
\label{fig:GPCase-2-LowT}
\end{figure}

Now, we intend to study the effects to the isotropic parameter $d_{00}$ in the thermodynamic functions of the system. As we did before, we present the particle number in Fig. \ref{fig:GPCase-2-lowTN}. $\mathcal{N}(T,\xi)$ is a concave function of temperature and its local minimum is bounded from below by the value $\xi=0.9$. For high temperature regime, we observe the parameter $\xi$ becoming negligible in the function $\mathcal{N}(T,\xi)$ when $T>1.4~\mathrm{eV}$. The minimum value is shifted to the left until it reaches $T\approx0.63~\mathrm{eV}$ for $\xi=0.9$. After this value, the minimum moves to the right again. We can use these information to control the number of particle inside the ring. Such a feature has many applications in nanotechnology \cite{chakraborty1994electron,warburton2000optical,fuhrer2001energy}.

To the internal energy, see Fig. \ref{fig:GPCase-2-lowTU}. We observe that, for high temperature regime, the parameter $\xi$ become negligible and the function $\mathcal{U}(T,\xi)$ does not depend on $\xi$. For low temperature regime, mainly in the range $0<T<0.6~\mathrm{eV}$, we observe an splitting in the energy. This means that the function $\mathcal{U}(T,\xi)$ is sensible to different values of $\xi$. Here, the energy is bounded from below, and, for $\xi=0.9$, and, for $\xi>0.9$, the internal energy increases. For $T=0$, we find that $-1.5~\mathrm{eV}<\mathcal{U}(T=0,\xi)<-1.0~\mathrm{eV}$. The freedom in having to choose parameter $\xi$ can also be of great importance in several applications, e.g., it controls the internal motions (kinetic energy) of electrons.

We finish this subsection by looking at the behavior of the heat capacity, which is shown in Fig. \ref{fig:GPCase-2-lowTC}. In agreement with what happens to the all thermodynamic functions studied in this manuscript, the heat capacity also becomes $\xi$ independent if $T$ reaches high values. There still exists a formation of a local maximum when we increase the values of $\xi$. More so, the maximum value reached by this local maximum is bounded from $\xi=0.9$ and, for $\xi>0.9$. Remarkably, we can utilize conveniently $\xi$ in order to maximize of even minimize the heat capacity of a quantum ring.

\subsection{Spin currents}

Here, we investigate the behavior of spin current when temperature is modified. To study it, we need to evaluate the mean value $\mathcal{J}_{\varphi }^{z}$ with respect to the corresponding distribution of probability. In this sense, the average value of the spin current is given by%
\begin{equation}
\left\langle \mathcal{J}_{\varphi }^{z}\right\rangle =\sum_{n=0}^{\infty
}\sum_{\left\{ s,\lambda \right\} =\left\{ \pm \right\} }\frac{1}{4\mathrm{m}r_{0}}%
\frac{\left( 2n\cos \theta -1\right) }{\exp \left[ \beta \left(
E_{n,s}^{\lambda }-\mu \right) \right] +1}.
\end{equation}%

\begin{figure}[tbh]
\centering
  \subfloat[Spin current for anisotropic coefficient $d_{ij}$]{\includegraphics[width=8cm,height=5cm]{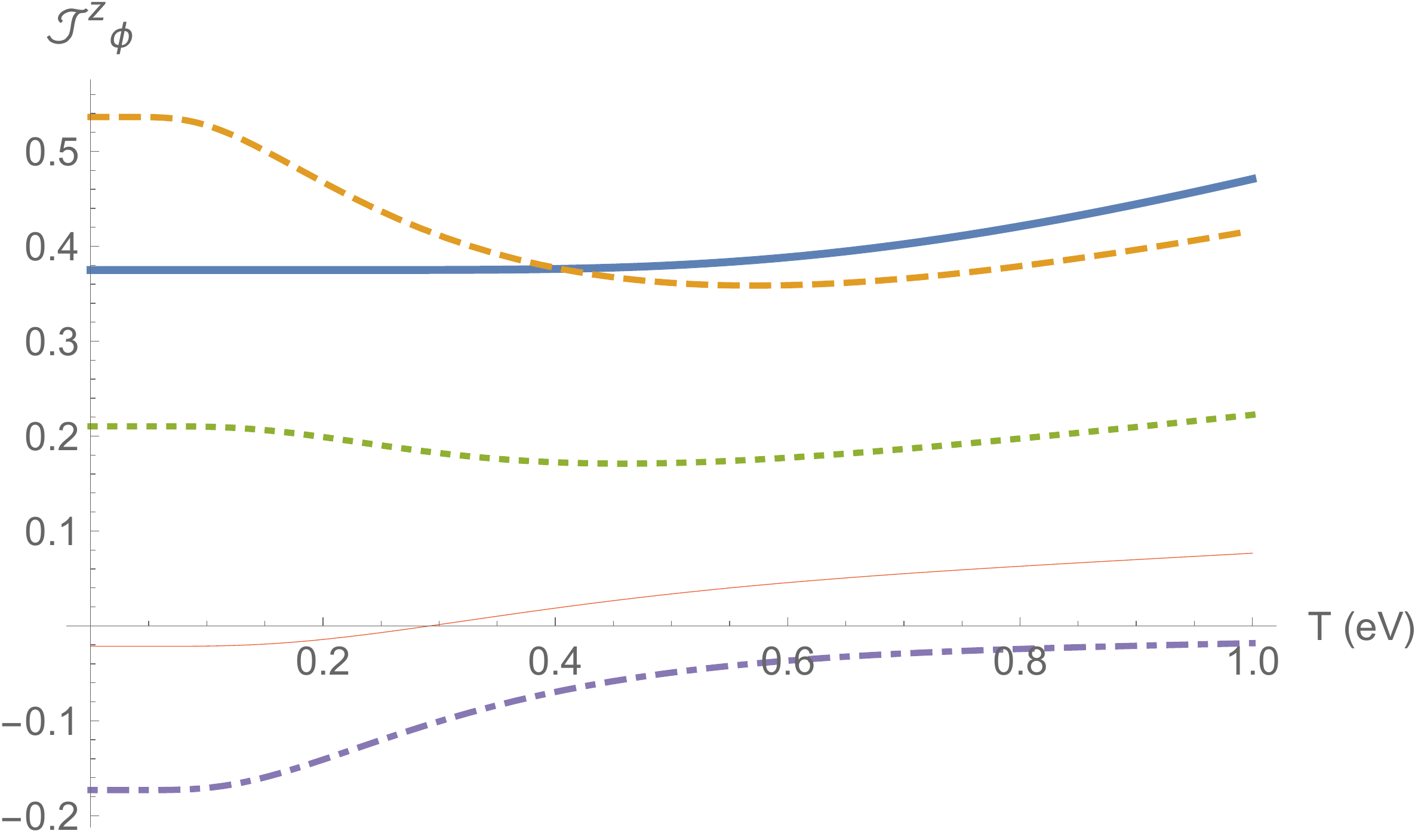}
  \label{fig:CCase2-f1}}
  \subfloat[Spin current for isotropic coefficient $d_{00}$]{\includegraphics[width=8cm,height=5cm]{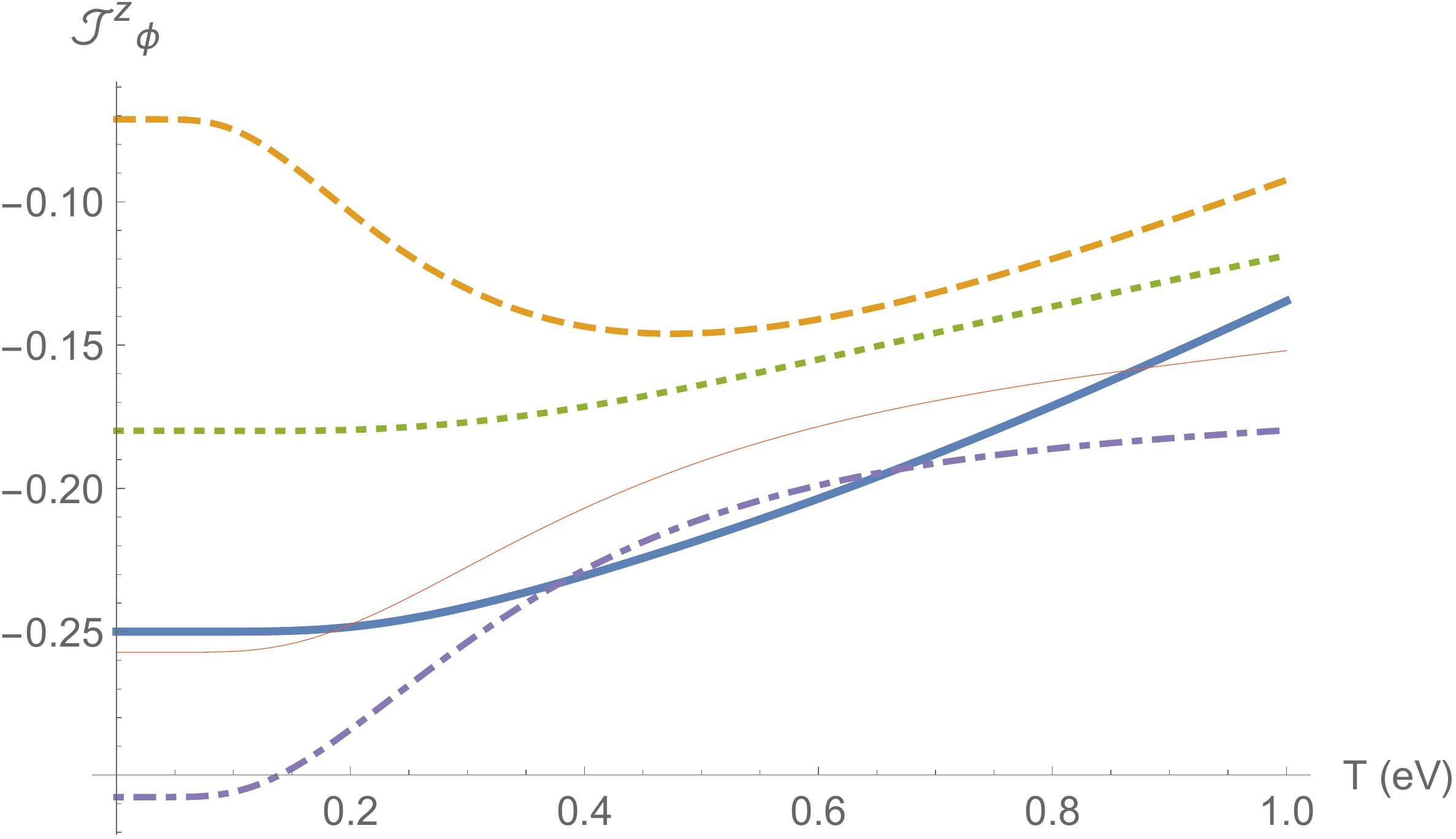}
  \label{fig:CCase2-f2}}
\caption{The spin currents results for the grand canonical approach are presented above, where the labels (thick, blue, $\{\xi,\xi_{00}\}=0.0$), (dashed, orange, $\{\xi,\xi_{00}\}=0.3$), (dotted, green, $\{\xi,\xi_{00}\}=0.6$), (thin, red, $\{\xi,\xi_{00}\}=0.9$), (dotdashed, purple, $\{\xi,\xi_{00}\}=1.2$) identify the values of $\xi$.}
\label{fig:GPCaseJ}
\end{figure}

In order to analyze this expression numerically, we display the plots in Fig. \ref{fig:GPCaseJ} for both cases, namely, $d_{ij}$ and $d_{00}$. $\mathcal{J}_{\varphi }^{z}(T,\xi)$ is a decreasing function of $\xi$ up to $\xi=0$ configuration. We also see that, for $\xi=1.2$, electrons move in an opposite direction when compared to values in the range $0\leq\xi\leq0.6$. As mentioned in Ref. \cite{Scurrent}, the system can distinguish a clockwise and a counterclockwise moving current.  
We see that temperature has more influence under the counterclockwise motion for the anisotropic case. On the other hand, for the isotropic case, we observe that for $\xi=1.2$ there is a turning point in the motion at $T=0.15~\mathrm{eV}$ and electrons start moving clockwise.  

\begin{figure}[tbh]
\centering
  \subfloat[Internal energy for coefficient $d_{ij}$]{\includegraphics[width=8cm,height=5cm]{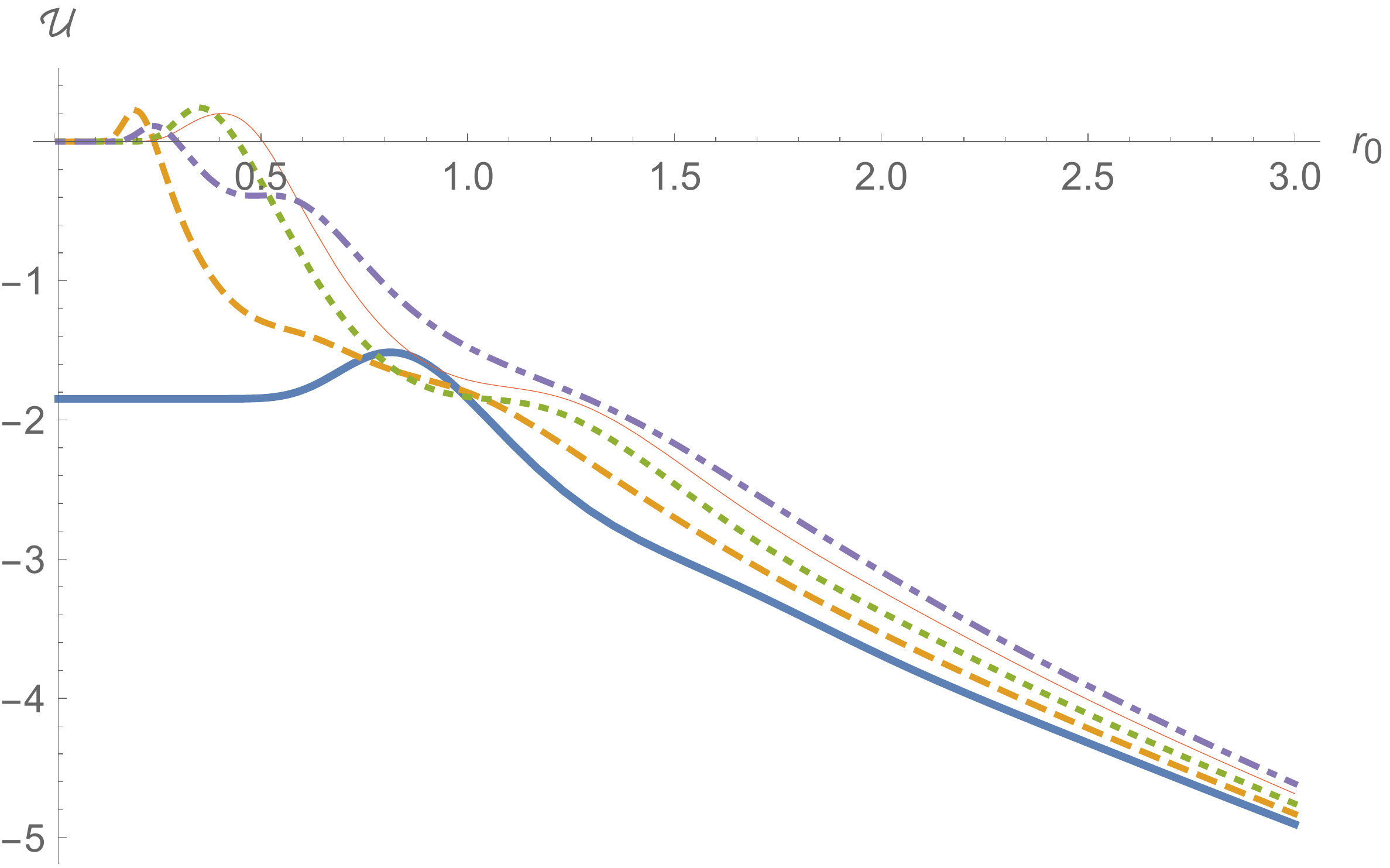}
  \label{fig:UROCase1}}
  \subfloat[Internal energy for coefficient $d_{00}$]{\includegraphics[width=8cm,height=5cm]{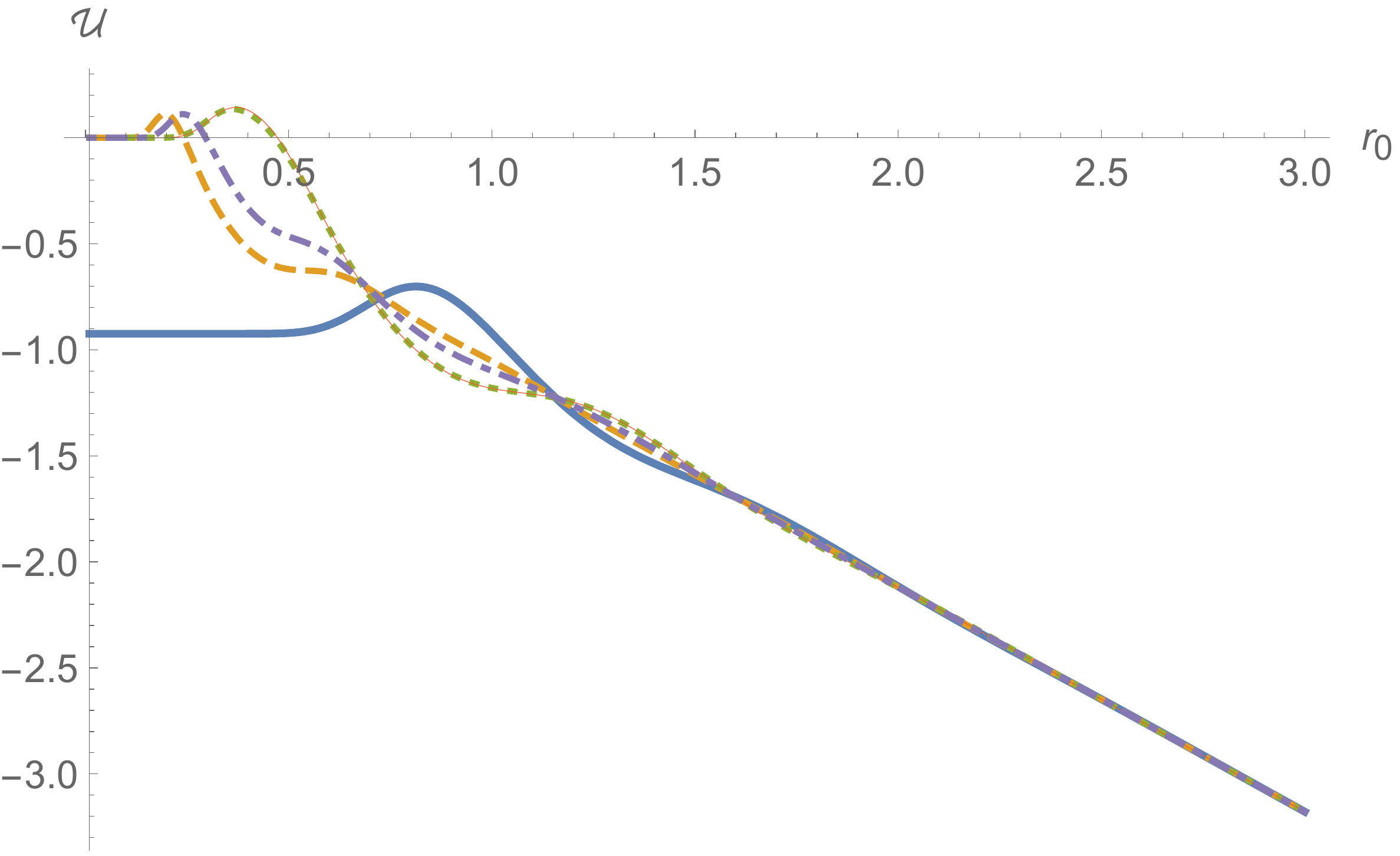}
  \label{fig:UR0Case2}}
\caption{The internal energy for the grand canonical approach are presented above now as a function of $r_{0}$, where the labels (thick, blue, $\{\xi,\xi_{00}\}=0.0$), (dashed, orange, $\{\xi,\xi_{00}\}=0.3$), (dotted, green, $\{\xi,\xi_{00}\}=0.6$), (thin, red, $\{\xi,\xi_{00}\}=0.9$), (dotdashed, purple, $\{\xi,\xi_{00}\}=1.2$) identify the values of $\xi$ for $T = 0.4~\mathrm{eV}$.}
\label{fig:UR0}
\end{figure}

Let us now examine how the internal energy, heat capacity, and spin currents change as the radius $r_{0}$ varies. In the upcoming plots, we compare two cases with a fixed temperature and different values for the Lorentz--violating coefficients. We begin by analyzing the internal energy, which is displayed in Fig. \ref{fig:UR0}. It is interesting to note that in both cases, the internal energy reaches a local maximum before decreasing as the radius increases. Furthermore, for larger values of $r_{0}$, the internal energy tends to stabilize and assume the same value. This is expected since the Lorentz violating coefficients become increasingly suppressed. Comparing the two cases, we see that the isotropic configuration produces a system with higher internal energy than the anisotropic one.  

\begin{figure}[tbh]
\centering
  \subfloat[Heat capacity for coefficient $d_{ij}$]{\includegraphics[width=8cm,height=5cm]{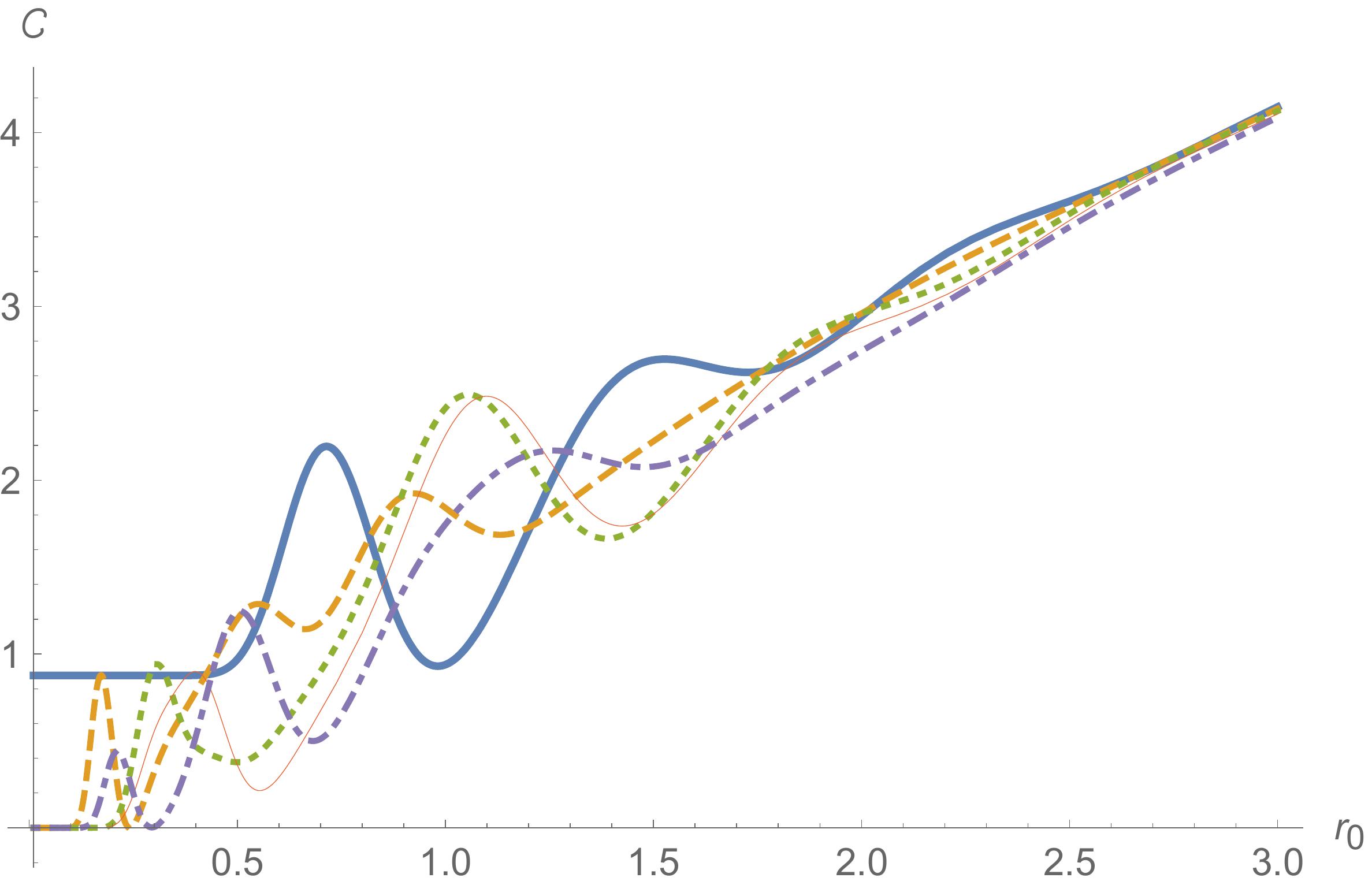}
  \label{fig:CROCase1}}
  \subfloat[Heat capacity for coefficient $d_{00}$]{\includegraphics[width=8cm,height=5cm]{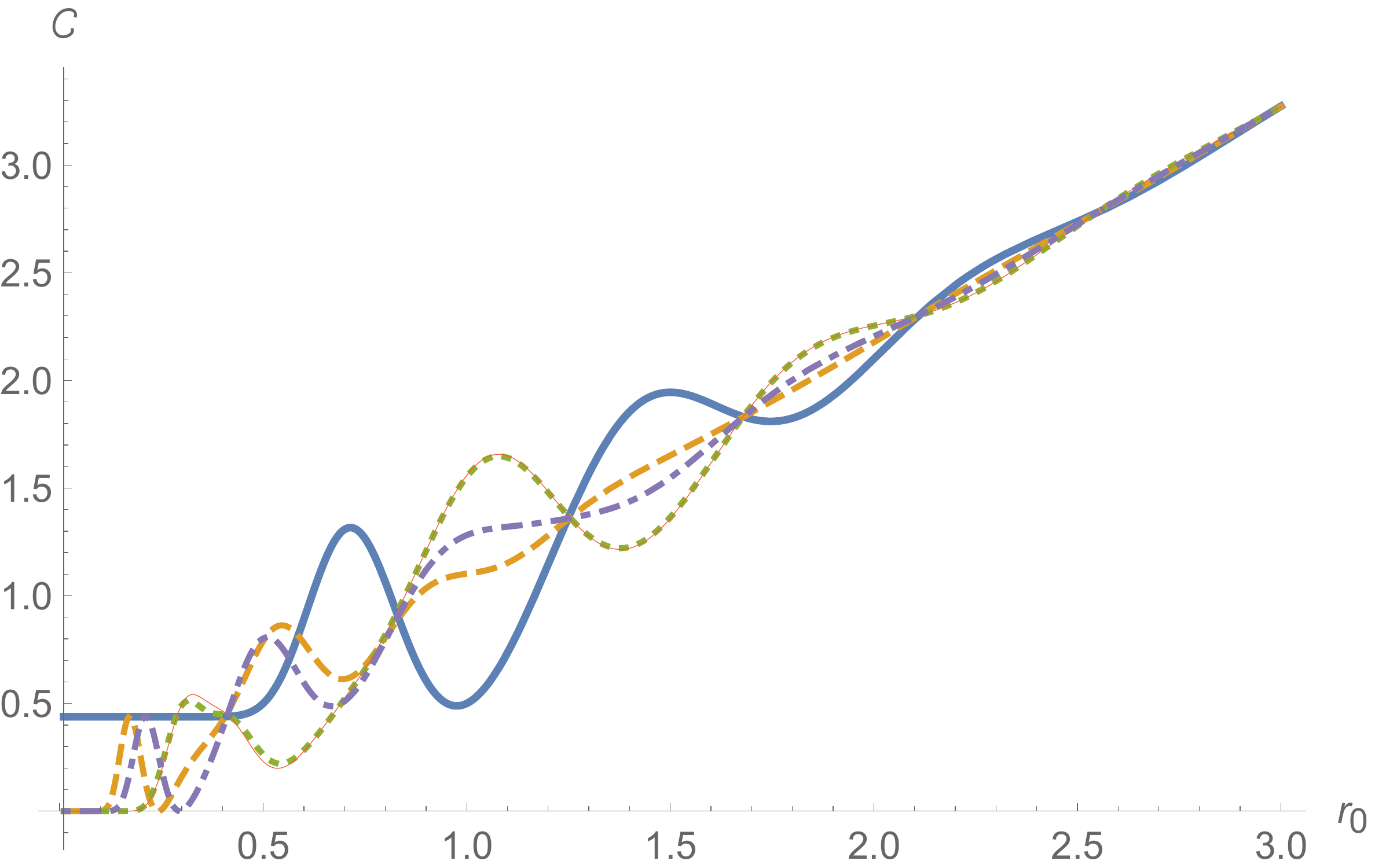}
  \label{fig:CR0Case2}}
\caption{The Heat capacity for the grand canonical approach are presented above now as a function of $r_{0}$, where the labels (thick, blue, $\{\xi,\xi_{00}\}=0.0$), (dashed, orange, $\{\xi,\xi_{00}\}=0.3$), (dotted, green, $\{\xi,\xi_{00}\}=0.6$), (thin, red, $\{\xi,\xi_{00}\}=0.9$), (dotdashed, purple, $\{\xi,\xi_{00}\}=1.2$) identify the values of $\xi$ for $T = 0.4~\mathrm{eV}$.}
\label{fig:CR0}
\end{figure}

In contrast to the internal energy, the heat capacity exhibits a periodic behavior as the radius increases, as shown in Fig. \ref{fig:CR0}. For larger values of $r_{0}$, the heat capacity tends to stabilize and assume the same value, which is expected due to the diminishing effects of Lorentz violation. Interestingly, the results for the anisotropic configuration produces a system with a larger heat capacity.

\begin{figure}[tbh]
\centering
  \subfloat[Spin current for coefficient $d_{ij}$]{\includegraphics[width=8cm,height=5cm]{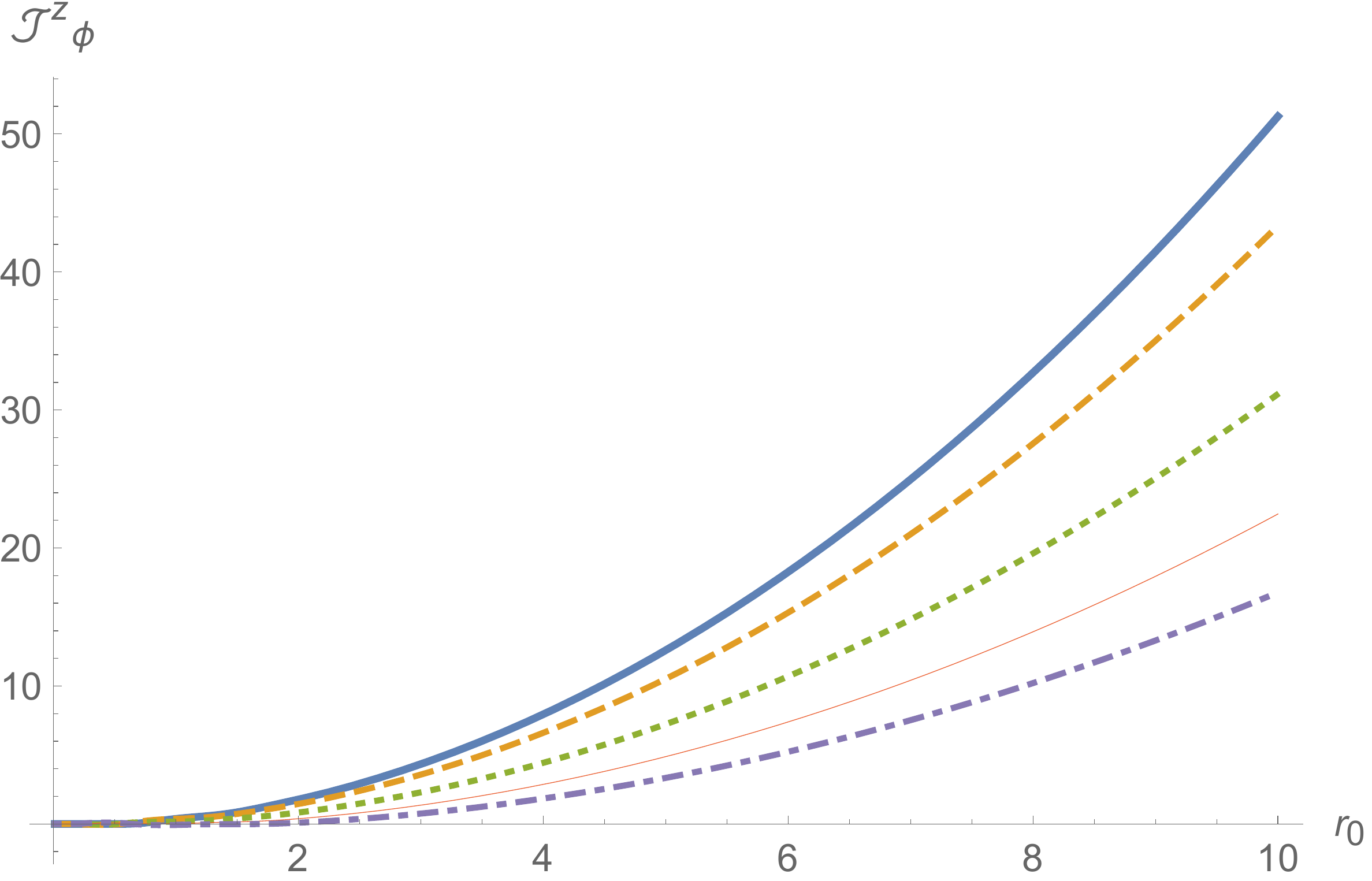}
  \label{fig:SROCase1}}
  \subfloat[Spin current for coefficient $d_{00}$]{\includegraphics[width=8cm,height=5cm]{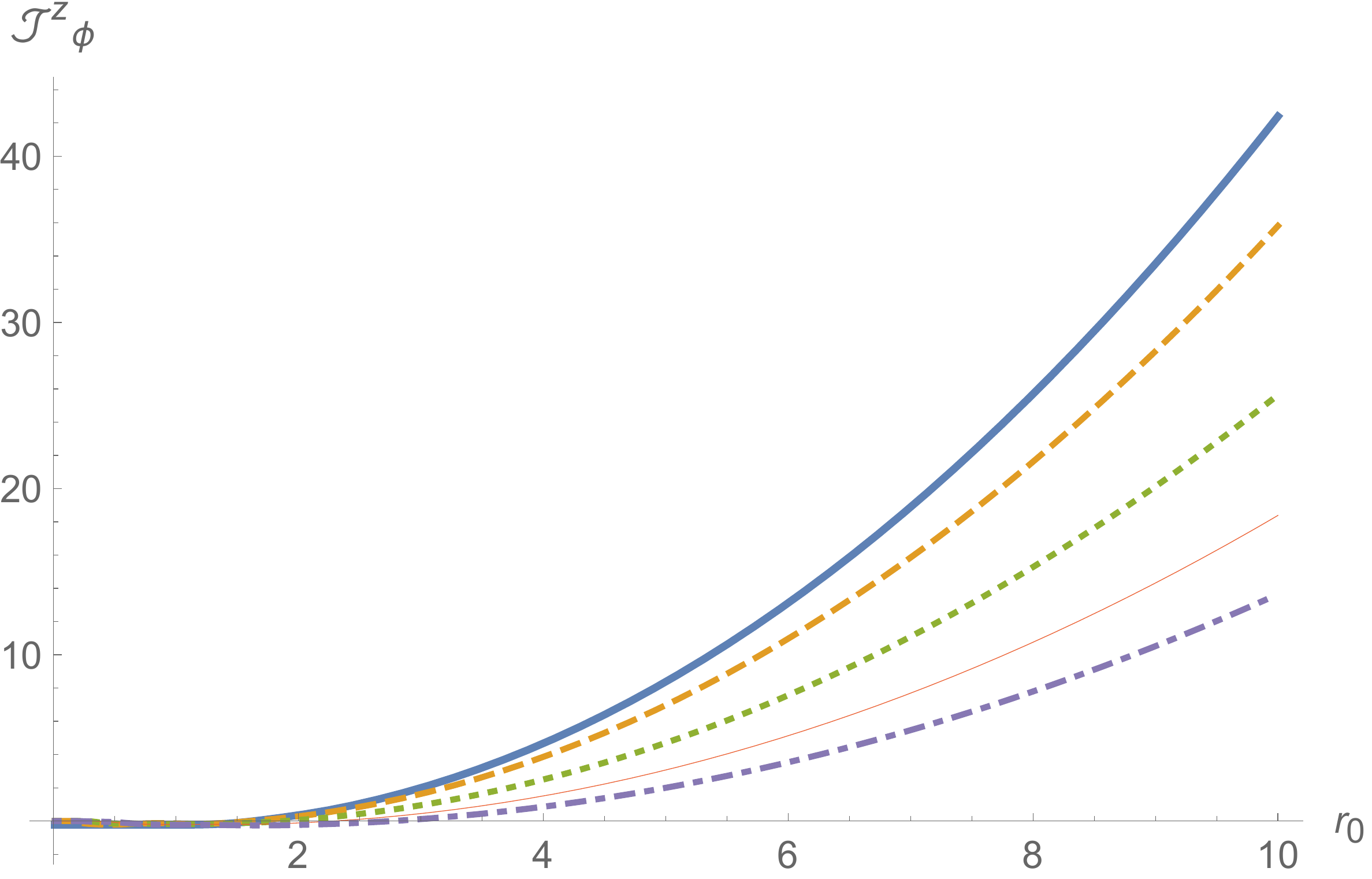}
  \label{fig:SR0Case2}}
\caption{The spin current for the grand canonical approach are presented above now as a function of $r_{0}$, where the labels (thick, blue, $\{\xi,\xi_{00}\}=0.0$), (dashed, orange, $\{\xi,\xi_{00}\}=0.3$), (dotted, green, $\{\xi,\xi_{00}\}=0.6$), (thin, red, $\{\xi,\xi_{00}\}=0.9$), (dotdashed, purple, $\{\xi,\xi_{00}\}=1.2$) identify the values of $\xi$ for $T = 0.4~\mathrm{eV}$.}
\label{fig:SR0}
\end{figure}

Finally, Fig. \ref{fig:SR0} illustrates the behavior of the spin current. It is evident that the magnitude of the spin current increases as the radius grows larger. Furthermore, a comparison of the two cases reveals that the anisotropic configuration produces a larger spin current.

\section{Conclusions}

This manuscript had the purpose of calculating the thermodynamic properties of a quantum ring in the context of Lorentz violation. To do so, we fundamentally used the ensemble approach.

Within the canonical ensemble, we calculated the partition functions for both operators $d_{ij}$ and $d_{00}$. With these ones, we were able to calculate particle number,
entropy, mean energy, heat capacity, and spin currents. In addition, we observed that parameter $\xi$ played a major role at low temperature regime. In other words, the thermodynamic properties of a quantum ring turned out to be sensible to different values of $\xi$. Such a feature might probably be useful in the future nanotechnology in the case of the existence of Lorentz violation.

Another interesting property was related to the spin currents. In both scenarios, we saw that electrons could move clockwise or counterclockwise by choosing particular ranges of temperature and $\xi$. We noticed an inversion in their motion due to the change on Lorentz--violating parameters. When we compared the usual Rashba-- and Dresselhauss--interactions, we realized that the controlling coefficients played the role of an external magnetic field.

As a future perspective, we could also examine the aforementioned properties, replacing electrons by other spin half particles or nanoparticles. These systems are now under consideration and they will appear in a forthcoming paper.

\section*{Acknowledgments}
\hspace{0.5cm}

The authors also express their gratitude to FAPEMA, CNPq and CAPES (Brazilian research agencies) for invaluable financial support. In particular, J.A.A.S. Reis is supported by FAPEMA BPD-08734/22, L. Lisboa-Santos is supported by FAPEMA BPD-11962/22 and A. A. Araújo Filho is supported by Conselho Nacional de Desenvolvimento Cientíıfico e Tecnológico (CNPq) -- 200486/2022-5. More so, the authors would like to thank the referees for their useful comments and suggestions given to us.

\section{Data Availability Statement}

Data Availability Statement: No Data associated in the manuscript


\bibliographystyle{ieeetr}
\bibliography{main}

\end{document}